\documentclass[useAMS,usenatbib]{mn2e}

\pdfoutput=1

\usepackage{ifpdf}
\usepackage{amssymb,amsmath}
\usepackage{graphics}
\usepackage{subfigure}
\usepackage[dvips]{graphicx}
\usepackage{epstopdf}
\usepackage{textcomp}
\usepackage[english]{babel}
\usepackage{color}

\title{SPHGal: Smoothed Particle Hydrodynamics with improved accuracy for galaxy simulations}

\author[Hu et al.]
{Chia-Yu Hu$^{1}$, Thorsten Naab$^{1}$, Stefanie Walch$^{2,1}$, Benjamin P. Moster$^{3,1}$, 
\newauthor Ludwig Oser$^{4}$\\
$^{1}$Max-Planck-Institut f\"ur Astrophysik,
Karl-Schwarzschild Strasse 1, D-85740 Garching, Germany\\
$^{2}$Physikalisches Institut der Universit\"at zu K\"oln, 
Z\"ulpicher Strasse 77, D-50937 K\"oln, Germany\\
$^{3}$Institute of Astronomy, Madingley Road, Cambridge CB3 0HA, UK\\
$^{4}$Department of Astronomy, Columbia University, New York, NY 10027, USA
}

\begin{document}
\maketitle

\begin{abstract}
We present the smoothed-particle hydrodynamics implementation SPHGal, which combines
some recently proposed improvements in GADGET. This includes a pressure-entropy
formulation with a Wendland kernel, a higher order estimate of velocity gradients, a
modified artificial viscosity switch with a modified strong limiter, and artificial
conduction of thermal energy. With a series of idealized hydrodynamic tests we show
that the pressure-entropy formulation is ideal for resolving fluid mixing at contact
discontinuities but performs conspicuously worse at strong shocks due to the large
entropy discontinuities. Including artificial conduction at shocks greatly improves
the results. In simulations of Milky Way like disk galaxies a feedback-induced
instability develops if too much artificial viscosity is introduced. Our modified
artificial viscosity scheme prevents this instability and shows efficient shock
capturing capability. We also investigate the star formation rate and the galactic
outflow. The star formation rates vary slightly for different SPH schemes while the
mass loading is sensitive to the SPH scheme and significantly reduced in our favored
implementation.  We compare the accretion behavior of the hot halo gas. The
formation of cold blobs, an artifact of simple SPH implementations, can be
eliminated efficiently with proper fluid mixing, either by conduction and/or by
using a pressure-entropy formulation.
\end{abstract}

\begin{keywords}
methods: numerical,  galaxies: ISM, galaxies: intergalactic medium, galaxies:
evolution, galaxies: spiral
\vspace{-1.0cm}
\end{keywords}

\section{Introduction}
Smoothed-particle hydrodynamics (SPH) is a numerical method for
solving fluid equations in a Lagrangian fashion
\citep{1977AJ.....82.1013L, 1977MNRAS.181..375G}. It has found a
widespread use in a variety of astrophysical problems
(e.g. \citealp{2010ARA&A..48..391S,2012JCoPh.231..759P}). The success
of SPH lies in the exact conservation of physical properties and the
adaptive resolution that traces mass. In addition, the Lagrangian
nature ensures Galilean invariance by construction. Practically, the
method is also straightforward to be incorporated with tree-based
gravity solvers and is intuitive for including sub-resolution
physics. 

However, recent studies (e.g. \citealp{2007MNRAS.380..963A,
  2010MNRAS.401..791S}) have shown that standard SPH has serious
difficulties to properly model fluid mixing. Its accuracy is therefore
compromised when modeling the interstellar medium
\citep{2011MNRAS.415..271H} and galaxy formation
\citep{2012MNRAS.424.2999S, 2013MNRAS.429.3353N}. The problem
originates from a numerical artifact at contact discontinuities (often 
referred to as the spurious 'surface tension'). One solution proposed
by \citet{2008JCoPh.22710040P} is to include artificial conduction to
alleviate such artifacts, though with the potential issue of being
overly diffusive. Therefore, several conduction switches have been
suggested to reduce unwanted conduction away from the entropy
discontinuities \citep{2012MNRAS.422.3037R,
  2012A&A...546A..45V}. An alternative solution to the problem is to
modify the definition of density so that the pressure is smoothed by
construction \citep{2001MNRAS.323..743R, 2010MNRAS.405.1513R, 2013ApJ...768...44S,
  2013MNRAS.428.2840H}. To its advantage no extra dissipation is
needed, the entropy is still conserved and the contact discontinuities 
remain sharp. 

Another criticism of SPH is its slow convergence rate. SPH is second
order accurate only in the continuous limit. When the fluid is
discretized into SPH particles, there exists a zeroth order error in
the equation of motion \citep{2010MNRAS.405.1513R,
  2012JCoPh.231..759P}. This error can be reduced by using more
neighboring particles within the kernel.  However, the commonly used
cubic spline kernel is subject to the pairing instability when too
many neighboring particles are used \citep{2012JCoPh.231..759P}.  
\citet{2010MNRAS.405.1513R} introduced a new kernel function without
an inflection point to prevent the pairing
instability. \citet{2012MNRAS.425.1068D} proposed to use the Wendland $C^4$ 
kernel and demonstrated its stability despite having an inflection
point.  These studies demonstrate a significantly improved convergence
rate compared to the standard SPH with only a minor impact on the
computational efficiency.  

With these recent developments, many authors have updated the existing
'problematic' implementations to improve the accuracy and ensure the
scientific credibility of their simulations. \citet{2013arXiv1307.0668P}
presented simulations of the formation of an idealized (non-radiative)
galaxy cluster using three different methods: the traditional SPH,
their improved implementation (SPHS) and an adaptive mesh refinement
(AMR) code (see also \citet{2013MNRAS.432..711H} for a recent AMR/SPH
comparison). They find that SPHS and AMR are in excellent agreement
while the traditional SPH shows a different behavior. 
On the other hand, \citet{2013arXiv1311.2073H} presented cosmological
simulations including more complicated physical processes and found
very little difference between their improved SPH implementation and
the traditional SPH. \citet{2013arXiv1312.1391F} simulated the
accretion of supermassive black holes at high redshifts and found that
their improved SPH gives rise to slightly faster black hole growth,
while the star formation is unaffected.  

In this paper, we incorporate the most important improvements into the
hydrodynamic code GADGET-3 \citep{2005MNRAS.364.1105S} and test
different implementations with idealized hydrodynamic tests. We
use the pressure-entropy formulation presented in
\citet{2013MNRAS.428.2840H} and the Wendland $C^4$ kernel with 200
neighboring particles.  The artificial viscosity scheme we adopt resembles \citet{2010MNRAS.408..669C} but with a slight modification
in its functional form. We also include artificial conduction similar
to \citet{2012MNRAS.422.3037R}. However, the major role of conduction
in our case is to help capture shocks. We show that the
pressure-entropy formulation performs poorly in the Sedov explosion 
test without artificial conduction. At contact discontinuities,
however, the pressure-entropy formulation alone is able to properly
model the fluid mixing. We therefore suppress the artificial
conduction in shear flows.

We also investigate the properties of different SPH schemes in a more
complex three dimensional simulation of isolated disk galaxy models with
gas, star formation and supernova feedback.
Standard SPH implementations can result in the formation of artificial
large, kpc-sized, holes which are then sheared apart. If too much
viscosity is introduced the gaseous disks become unstable and the
simulations fail dramatically.  Fortunately, such instabilities can be
avoided with a slight modification of the viscosity limiter.  

We present a modified SPH version, which we term SPHGal, 
that passes the Gresho, Sod shock tube,
Sedov explosion, 'square', Keplerian ring, Kelvin-Helmholtz and 'blob'
test and also performs well in more realistic galaxy
simulations. The strengths and limitations of this implementation are
discussed in detail.

This paper is organized as follows: in Section \ref{sec:method} we
present the details of our SPH implementation. The results of the
idealized hydrodynamic tests are shown in Section \ref{sec:hydrotest}
with detailed discussions of the pros and cons of different
implementations. In Section \ref{sec:disk} we present simulations of
isolated disk galaxy models and show the properties 
of the gaseous disk, the star formation rate and galactic outflow, and the accretion behaviors. We summarize and discuss
our work in Section \ref{sec:conclusion}.

\section{Hydrodynamic Method}\label{sec:method}

\subsection{Improving the convergence rate}
In SPH, the pressure force in the equation of motion is not guaranteed
to vanish in a medium of constant pressure, which is referred to as
the "E0 error" in \citet{2010MNRAS.405.1513R}.  This residual force
vanishes only when SPH particles are distributed regularly within the
kernel. The convergence scaling of SPH is therefore, in general,
worse than $O(h^2)$. Although this error can be factored out to obtain
a locally more precise form of the pressure gradient, the inevitable
trade-off is the violation of exact momentum conservation and losing
the capability of particle re-ordering \citep{2012JCoPh.231..759P},
making it less favorable in practice. 
One straightforward way of reducing the E0 error is to increase the
particle number in the kernel so that the integration accuracy is
improved. However, the commonly used cubic spline kernel is subject to
the pairing instability when using too many neighboring particles
\citep{2012JCoPh.231..759P}. Alternative kernel functions immune to 
pairing instability have been proposed
\citep{2010MNRAS.405.1513R,2012MNRAS.425.1068D}. Here we adopt the
Wendland $C^4$ kernel as in \citet{2012MNRAS.425.1068D} and use 200
neighboring particles as our default setup.

\subsection{Pressure-entropy formulation}
It has been widely recognized that the standard SPH does not properly
model fluid mixing at contact discontinuities
\citep{2001MNRAS.323..743R,2007MNRAS.380..963A}. The problem is that
while the density is smoothed on a kernel scale at the boundaries, the
entropy remains sharply discontinuous, leading to a so-called
"pressure blip" which acts as a spurious surface tension, suppressing
fluid instabilities. More than ten years ago
\citet{2001MNRAS.323..743R} derived an alternative density estimate
which can avoid this numerical artifact. 
\citet{2010MNRAS.405.1513R} proposed a generalized discretization of
the Euler equation where the density estimate in \citet{2001MNRAS.323..743R} 
is a special case.
\citet{2013ApJ...768...44S}
explored a similar idea and derived a density-independent SPH
formulation by choosing a different volume element.
\citet{2013MNRAS.428.2840H} took into account the variation of the
smoothing length (the "grad-h" term) based on a Lagrangian approach,
making the new formulation exactly conservative,
which is important for modeling shocks.

We adopt the pressure-entropy (PE) formulation derived in
\citet{2013MNRAS.428.2840H}, where pressure and entropy are the
primary variables. The pressure is estimated by
\begin{equation}
	\widehat{P}_{i} = \left[ \sum_{j=1}^{N} m_j A_j^{1/\gamma}
          W_{ij}(h_j) \right] ^{\gamma}, 
\end{equation}
where $N$ is the number of neighboring particles in the kernel, $m_j$
is the particle mass, $W_{ij}$ is the smoothing kernel, $h_j$ is the
smoothing length, $A_j$ is the entropy function and $\gamma$ is the
polytropic index such that $P = A \rho^{\gamma}$. The equation of
motion is 
\begin{eqnarray}\label{eq:eom}
\begin{aligned}
	\frac{{\rm d}{\bf v}_i}{{\rm d}t} = &
	 - \sum_{j=1}^{N} m_j (A_i A_j)^{1/\gamma} \\
	& ~~~\times \left[ 
	\frac{ f_{ij} \widehat{P}_i}{\widehat{P}_i^{2/\gamma}} \nabla_i W_{ij}(h_i) + 
	\frac{ f_{ji} \widehat{P}_j}{\widehat{P}_j^{2/\gamma}} \nabla_i W_{ij}(h_j)
	\right], 
\end{aligned}
\end{eqnarray}
where 
\begin{equation}
	f_{ij} = 1 - \left( \frac{h_i}{3 A_j^{1/\gamma} m_j \widehat{n}_i }
	\frac{\partial \widehat{P}_i^{1/\gamma}}{\partial h_i}  \right)
	\left[ 1 + \frac{h_i}{3\widehat{n}_i}\frac{\partial \widehat{n}_i}{\partial h_i}
	\right]^{-1}
\end{equation}
is the correction term for variable smoothing lengths and
\begin{equation}
	\widehat{n}_i = \sum_{j=1}^{N} W_{ij}
\end{equation}
is the number density estimate. The entropy is given by the initial
conditions and requires no evolution in the dissipation-less
case. Other thermodynamic variables (e.g. density) can be derived from
the estimated pressure and entropy.

The advantage of such a formulation is that the pressure is smoothed
by construction and therefore has no spurious jump at contact
discontinuities. Fluid instabilities can thus develop without being
numerically suppressed, and the hot and cold regions of the fluid then
mix with each other within a few instability time-scales. However,
this formulation is not without its weaknesses. In many astrophysical
situations, especially at strong shocks, the variation in entropy from 
one particle to another is usually several orders of magnitude larger
than in density. A particle with high entropy would then have an
overwhelming weight even if it is located at the edge of the
kernel. The pressure estimate, in such cases, would be much noisier or
even biased compared to the standard density-entropy (DE) SPH
formulation. 
\textbf{
This can be seen most clearly by linear error analysis \citep{2010MNRAS.405.1513R}.
Assuming the smoothing length is constant,
Equation (\ref{eq:eom}) can be written as
\begin{eqnarray}
\begin{aligned}	
	\frac{{\rm d}{\bf v}_i}{{\rm d}t} = &	
	- \sum_{j=1}^{N} \frac{m_j}{\rho_i \rho_j} 
	\left[ 
	\frac{\rho_j}{\rho_i}\frac{A_j^{1/\gamma}}{A_i^{1/\gamma}} P_i + 
	\frac{\rho_i}{\rho_j}\frac{A_i^{1/\gamma}}{A_j^{1/\gamma}} P_j  
	\right]
    \nabla_i W_{ij}\\
    = &
    - \frac{P_i}{\rho_i} \sum_{j=1}^{N} \frac{m_j}{\rho_j} 
	\left[ 
	\frac{\rho_j}{\rho_i}\frac{A_j^{1/\gamma}}{A_i^{1/\gamma}}  + 
	\frac{\rho_i}{\rho_j}\frac{A_i^{1/\gamma}}{A_j^{1/\gamma}}   
	\right]
    \nabla_i W_{ij}\\
	& - \frac{\nabla P_i}{\rho_i}  
	\sum_{j=1}^{N} \frac{m_j}{\rho_j} 
	\frac{\rho_i}{\rho_j}\frac{A_i^{1/\gamma}}{A_j^{1/\gamma}}
	({\bf x}_j - {\bf x}_i)\otimes \nabla_i W_{ij} + O(h^2),
\end{aligned}
\end{eqnarray}
where we have used $P_j = P_i + \nabla P_i \cdot ({\bf x}_j - {\bf x}_i) + O(h^2)$.
Therefore,
for the equation of motion to be second order accurate,
the following two conditions have to be satisfied:
\begin{eqnarray}
\begin{aligned}	
	{\bf E}_i &\equiv \sum_{j=1}^{N} \frac{m_j}{\rho_j} 
	\left[ 
	\frac{\rho_j}{\rho_i}\frac{A_j^{1/\gamma}}{A_i^{1/\gamma}}  + 
	\frac{\rho_i}{\rho_j}\frac{A_i^{1/\gamma}}{A_j^{1/\gamma}}   
	\right]
    \nabla_i W_{ij} = {\bf 0},\\
    {\bf V}_i &\equiv
    \sum_{j=1}^{N} \frac{m_j}{\rho_j} 
	\frac{\rho_i}{\rho_j}\frac{A_i^{1/\gamma}}{A_j^{1/\gamma}}
	({\bf x}_j - {\bf x}_i)\otimes \nabla_i W_{ij} = {\bf I },
\end{aligned}
\end{eqnarray}
where ${\bf E}_i$ is the dominate error (E0 error).
At contact discontinuities,
$\rho_j A_j^{1/\gamma}\rho_i^{-1} A_i^{-1/\gamma} \approx 1$ across the interface,
which minimizes ${\bf E}_i$ as $\nabla_i W_{ij}$ is an odd function.
However,
at strong shocks,
$\rho_j A_j^{1/\gamma}\rho_i^{-1} A_i^{-1/\gamma}$ can deviate from unity by several orders of magnitude due to the large entropy jump.
Furthermore,
as both density and entropy are larger in the post-shock regions, their effects in ${\bf E}_i$ in general do not cancel out. 
}

\subsection{Artificial viscosity}\label{sec:AV}
We implement artificial viscosity (AV) as in
\citet{1997JCoPh.136..298M} and \citet{2005MNRAS.364.1105S}:  
\begin{equation}
	\left( \frac{{\rm d}{\bf v}_i}{{\rm d}t} \right)_{\rm{vis}} = 
	- \sum_{j=1}^{N} m_j \Pi_{ij} \nabla_i \overline{W}_{ij}, 
\end{equation}
where $\overline{W}_{ij}$ represents the arithmetic average of
${W}_{ij}(h_i)$ and ${W}_{ij}(h_j)$, and 
\begin{equation}\label{eq:vis_eq}
	\Pi_{ij} =
	\begin{cases}	
		 -\frac{1}{2} 
		\frac{{\bar\alpha}_{ij}v_{\rm sig}}{\bar{\rho}_{ij}} \omega_{ij}
		& \text{if } \omega_{ij} < 0 \\
		0& \text{otherwise },
	\end{cases}
\end{equation}
where $\omega_{ij} = {\bf v}_{ij}\cdot{\widehat{x}_{ij}}$ is the
approaching velocity of particle pairs, $ v_{\rm sig} = c_i + c_j -
3\omega_{ij}$ is the signal speed, $\bar{\rho}_{ij}$ is the arithmetic
average of $\rho_i$ and $\rho_j$, and $\bar{\alpha}_{ij}$ is the
arithmetic average of $\alpha_i$ and $\alpha_j$. The viscosity also
generates entropy at a rate 
\begin{equation}\label{eq:ent_AV}
	\frac{{\rm d} A_i}{\rm dt} = \frac{1}{2} \frac{\gamma - 1}{\rho_i^{\gamma-1}}
	\sum_{j=1}^{N} m_j \Pi_{ij}{\bf v}_{ij} \cdot \nabla_i \overline{W}_{ij} .
\end{equation}

This commonly adopted form of AV is devised to conserve
momentum. However, it does not distinguish between bulk and shear
viscosity since the viscous term $\Pi_{ij}$ involves only the relative
velocity of particle pairs, irrespective of the local velocity
gradient. This can lead to excessive viscosity especially in shear
flows, generating spurious angular momentum transport in a rotating
disk. A common reduction scheme is to include a limiter that
suppresses AV wherever the vorticity dominates over the velocity
divergence (\citealp{1989PhDT.......206B,1995JCoPh.121..357B}, see also
\citealp{2005MNRAS.364..753D,2009ApJS..184..326N}). More recently,
\citet{1997JCoPh.136...41M} proposed a variable viscosity coefficient
$\alpha_i$ for each SPH particle. The basic idea is that $\alpha_i$
should increase only when a converging flow is detected
($\nabla\cdot{\bf{v}} < 0$), and decays to a minimum value afterwards
in a few sound-crossing times. This "switch" efficiently suppresses 
unwanted viscosity away from shocks and has also been implemented in
other SPH codes (e.g \citealp{2009ApJS..184..298W}).   

\citet{2010MNRAS.408..669C} further improved this method in several
ways. They set $\alpha_i$ immediately to a desired value based on a
shock indicator to ensure that $\alpha_i$ rises rapidly enough
wherever needed. They also used the time derivative of velocity
divergence as a shock indicator to detect shocks in
advance. Finally, they implemented a more precise estimate of the
velocity gradient to prevent falsely triggered AV.  
Based on these principles, they proposed an AV scheme as follows:
a limiter similar to the the Balsara switch is defined as
\begin{equation}\label{eq:wl}
	\xi_i = \frac{|2 (1-R_i)^4 \nabla\cdot{\bf v}_i|^2}{|2 (1-R_i)^4 \nabla\cdot{\bf v}_i|^2 + {\bf S}_i^2 },
\end{equation}
where ${\bf S}_i = \sqrt{ {\bf S}_{\alpha\beta}{\bf S}_{\alpha\beta} }$ is the Frobenius norm of the shear tensor and
\begin{equation}
	R_i = \frac{1}{\rho_i} \sum_j {\rm sign} (\nabla\cdot{\bf v}) m_j W_{ij}.
\end{equation}
The shock indicator is defined as
\begin{equation}
	S_i = \xi_i {\rm max}(0,-\dot{\nabla}\cdot{\bf v}_i)
\end{equation}
and the target value of the viscosity coefficient is
\begin{equation} 
	\alpha_{{\rm tar},i} = \alpha_{\rm max} 
	\frac{h_i^2 S_i}{h_i^2 S_i + v_{{\rm dec},i}^2},
\end{equation}
where $v_{{\rm dec},i}$ is the decay speed
\begin{equation}
	v_{{\rm dec},i} = {\rm max}_{{\bf x}_{ij} \leq h_i} 
	(\bar{c}_{ij} - {\rm min}(0, \bf{v}_{ij}\cdot{\widehat{x}_{ij}})).
\end{equation}
The true viscosity coefficient of each particle evolves as
\begin{equation}
	\alpha_i = 
	\begin{cases}
		\alpha_{{\rm tar},i} & \text{if } \alpha_i \leq \alpha_{{\rm tar},i} \\
		\left[ \alpha_{{\rm tar} ,i} + (\alpha_i - \alpha_{{\rm tar},i}) 
		\exp(-{\rm d}t / \tau_i) \right]
		& \text{if } \alpha_i > \alpha_{{\rm tar},i},
	\end{cases}
\end{equation}
where ${\rm d}t$ is the time step and $\tau_i = 10 h_i / v_{\rm dec}$
is the decay time. One characteristic of this scheme is that the
limiter $\xi_i$ in Equation (\ref{eq:wl}) puts a relatively stronger
weight on the velocity divergence than on the shear. Furthermore, the
target value of the viscosity coefficient $\alpha_{{\rm tar},i}$ may
approach the maximum value $\alpha_{\rm max}$ if $h_i^2 S_i \gg
v_{{\rm dec},i}^2$ even when the limiter $\xi_i$ is small. 
Therefore, in the case where both significant shocks and shear flows
are involved, the effect of the limiter is weakened. We will therefore
refer to this scheme as artificial viscosity with a 'weak limiter'.

We have modified the above functional form so that the effect of the
viscosity limiter is stronger. This is to avoid too much viscosity
when both shocks and shear flows are present . We will come back to this in
more detail in Section \ref{sec:holes} where we discuss the
three-dimensional modeling of a disk galaxy. We adopt a slightly
modified form of $\alpha_i$ while still following the same
principles. We define the target value of the viscosity coefficient as 
\begin{equation} \label{eq:vis_target}
	\alpha_{{\rm tar},i} = \alpha_{\rm max} 
	\frac{h_i^2 S_i}{h_i^2 S_i + c_i^2}, 
\end{equation}
where $S_i = {\rm max}(0,-\dot{\nabla}\cdot{\bf v}_i)$ is the shock
indicator. The true viscosity coefficient of each particle evolves as 
\begin{equation}
	\alpha_i = 
	\begin{cases}
		\xi_i~\alpha_{{\rm tar},i} & \text{if } \alpha_i \leq \alpha_{{\rm tar},i} \\
		\xi_i \left[ \alpha_{{\rm tar} ,i} + (\alpha_i - \alpha_{{\rm tar},i}) 
		\exp(-{\rm d}t / \tau_i) \right]
		& \text{if } \alpha_i > \alpha_{{\rm tar},i}, 
	\end{cases}
\end{equation}
where ${\rm d}t$ is the time step and $\tau_i = 10 h_i / v_{\rm sig}$
is the decay time with the decay speed  
\begin{equation}
	v_{\rm dec} = {\rm max}_{{\bf x}_{ij} \leq h_i} 
	(\bar{c}_{ij} - {\rm min}(0, \bf{v}_{ij}\cdot{\widehat{x}_{ij}})),
\end{equation}
and $\xi_i$ is a limiter similar to the Balsara switch but in a quadratic form
\begin{equation}\label{eq:limiter}
	\xi_i = \frac{|\nabla\cdot{\bf v}_i|^2}{|\nabla\cdot{\bf v}_i|^2 + |\nabla\times{\bf v}_i|^2 + 0.0001 (c_i/h_i)^2}.
\end{equation}

We use the same higher order velocity gradient estimate as
\citet{2010MNRAS.408..669C} to prevent falsely triggered AV.  
With a functional form as in Equation (\ref{eq:vis_target}) AV is
suppressed in a subsonic converging flow and rises up to a maximum
value $\alpha_{\rm max}$ when the converging flow becomes supersonic.
The major difference of this scheme from the weak-limiter scheme is
that the limiter $\xi_i$ is placed such that $\alpha_i \leq \xi_i
\alpha_{\rm max}$ always holds. In addition, the limiter in Equation
(\ref{eq:limiter}) puts equal weights to the velocity divergence and
the vorticity. 
We adopt this scheme as our fiducial AV scheme and refer to it as
artificial viscosity with a 'strong limiter'.

There is one subtlety in Equations (\ref{eq:vis_eq}) and
(\ref{eq:ent_AV}) regarding the choice of $\rho_i$. In the PE
formulation, the density can be estimated either by $\rho^e_i =
(\widehat{P}_i / A_i)^{1/\gamma}$ (the "entropy-weighted" density), or
by the traditional definition $\rho_i = \sum m_j W_{ij}$ (the
"mass-weighted" density). We find the latter gives more accurate
results in the case of strong shocks due to the large entropy
jumps. Therefore, we use the traditional estimate whenever we need the
density information. This includes radiative cooling, the conversion
between entropy and internal energy, and  the gradient estimate in
Equations (\ref{eq:diff_eq}) and (\ref{eq:Brookshaw}).

\subsection{Artificial conduction}\label{sec:ac}
We include the artificial conduction (AC) of thermal energy similar to
\citet{2012MNRAS.422.3037R}, which explicitly conserves the total
energy within the kernel: 
\begin{equation}\label{eq:diff_eq}
	\frac{{\rm d} u_i}{{\rm d}t} = 
	\sum_{j=1}^{N} {\bar\alpha}^d_{ij} v_{\rm sig} L^p_{ij} m_j \frac{u_i - u_j}{\bar{\rho}_{ij}} 
	\widehat{x}_{ij} \cdot \nabla_i \overline{W}_{ij}, 
\end{equation}
where $v_{\rm sig} = (c_i + c_j - 3\omega_{ij})$ and $L^p_{ij} = |P_i
- P_j|/(P_i + P_j)$ is the pressure limiter proposed by
\citet{2012MNRAS.422.3037R}.  
In the continuous limit, Equation (\ref{eq:diff_eq}) recovers the
thermal conduction equation ${\rm d} u/{\rm d}t = \eta \nabla^2 u$
with a thermal conductivity, $\eta = \alpha^d_{ij}
v_{\rm sig} L^p_{ij} {\rm x}_{ij} / 2$, scaling with the local
resolution ${\rm x}_{ij} \lesssim h$.

Although the artificial conduction may be interpreted as the turbulent
mixing at sub-resolution scales \citep{2008MNRAS.387..427W}, the
motivation here is purely numerical. As pointed out in
\citet{2008JCoPh.22710040P}, just like momentum is smoothed by AV, the
thermal energy (or entropy) should also be smoothed so that it remains
differentiable everywhere. In the DE formulation of SPH, the presence
of AC mitigates the problematic pressure blip at contact
discontinuities. In the PE formulation the pressure is smoothed at
contact discontinuities by construction, which seems to imply that the
AC becomes redundant. However, the AC is still desirable (perhaps even
necessary) in the PE formulation, not at contact discontinuities, but
at strong shocks where the entropy jumps tend to be several orders of
magnitude. The pressure estimate would be very noisy if there were no
conduction to smooth out these entropy jumps.

To reduce artificial conduction away from entropy discontinuities, we
use a conduction switch similar to the AV switch. We define an entropy
jump indicator as the Laplacian of thermal energy in the form of
\citet{1985PASAu...6..207B} 
\begin{equation}\label{eq:Brookshaw}
	\nabla^2 u_i = 2 \sum_{j=1}^{N} m_j \frac{u_i - u_j}{\rho_j} 
	\frac{\nabla_i W_{ij}}{|{\bf x}_{ij}|}
\end{equation}
and a target conduction coefficient 
\begin{equation}
	\alpha^d_{{\rm tar},i} = \alpha^d_{\rm max} 
	\frac{|\nabla^2 u_i|}{|\nabla^2 u_i| + u_i/h_i^2}.
\end{equation}
The true conduction coefficient evolves as
\begin{equation}
	\alpha^d_i = 
	\begin{cases}
		\xi_i~\alpha^d_{{\rm tar},i} &\text{if }\alpha^d_i \leq \alpha^d_{{\rm tar},i} \\
		\xi_i\left[ \alpha^d_{{\rm tar} ,i} + (\alpha^d_i - \alpha^d_{{\rm tar},i}) 
		\exp(-{\rm d}t / \tau^d_i) \right]
		& \text{if } \alpha^d_i > \alpha^d_{{\rm tar},i}, 
	\end{cases}
\end{equation}
where $\tau^d_i = 2 h_i / v_{\rm dec}$ is the decay time and $\xi_i$
is the limiter defined in Equation (\ref{eq:limiter}) which suppresses 
the AC at shear flows. The limiter $\xi_i$ here also guarantees that
there will be no AC in a self-gravitating system under hydrostatic
equilibrium as $\nabla \cdot {\rm v}$ = 0 in such case.

\subsection{Timestep limiter}
As \citet{2009ApJ...697L..99S} pointed out, when modeling strong
shocks, the adaptive time step scheme in SPH requires an additional
constraint: the neighboring particles should have similar time steps
comparable within a factor of a few. We set the factor to
be 4 as our default choice. On top of that,
\citet{2012MNRAS.419..465D} further pointed out that when feedback
energy (thermal or kinetic alike) is injected, an inactive particle
should shorten its time step and become active immediately. This is to
ensure that ambient particles react to the explosion correctly and do
not remain inactive when a shock is approaching. 
We activate the inactive particles right after they receive feedback energy.

\section{Hydrodynamic Tests}\label{sec:hydrotest} 
\subsection{Naming convention}
We investigate different SPH schemes and their performance in the test
problems. For simplicity, we assign an acronym to a specific SPH
scheme (e.g. 'pe-avsl-ac', our fiducial scheme). The first segment 
indicates the adopted SPH formulation: 'pe' and 'de' are the
pressure-entropy formulation and density-entropy formulation,
respectively. The second segment indicates the AV scheme.  For
example, 'avB' is a constant AV with a Balsara switch, 'avwl' is
AV with a weak-limiter, and 'avsl' is AV with a strong-limiter (see
Section \ref{sec:AV}).  If AC is included, we add another segment
'ac'. Finally, 'erho' uses the entropy-weighted density in the
dissipation terms (AV and AC), and 'lvg' indicates the use of the
lower order velocity gradient estimator. Table \ref{table:name}
summarizes all the SPH schemes investigated in this paper. The maximum
value of the AV (and AC if applicable) coefficient is set to unity in
all cases. We also set a minimum value for the AV coefficient to 0.1
in order to maintain particle order. Unless otherwise specified, we
use the Wendland $C^4$ kernel with 200 neighboring particles. 

\begin{table*}
\caption{Naming convention of different SPH schemes. The maximum value
  of the AV (and AC if applicable) coefficient is set to be 1 in all
  cases. We also set a minimum value for the AV coefficient of 0.1 to
  retain the particle order. Our favored and fiducial SPH scheme is
  'pe-avsl-ac'.}  
\label{table:name}
\begin{tabular}{| l | c | c | c | c |}    
\hline\hline
Name   &  SPH formulation  & artificial viscosity (AV) & artificial conduction (AC) & comments\\
\hline
de-avB        &   density-entropy   &  const. AV with Balsara switch  &  no &  \\
de-avB-lvg    &   density-entropy   &  const. AV with Balsara switch  &  no &  lower order velocity gradients\\
de-avsl         &   density-entropy   &  strong limiter                     &  no &  \\
de-avsl-ac      &   density-entropy   &  strong limiter                     &  yes&  \\
de-avwl       &   density-entropy   &    weak limiter           &  no &  \\
de-avwl-ac    &   density-entropy   &  weak limiter            &  yes&  \\
pe-avB        &   pressure-entropy  &  const. AV with Balsara switch  &  no &  \\
pe-avsl         &   pressure-entropy  &  strong limiter                      &  no &  \\
pe-avsl-ac      &   pressure-entropy  &  strong limiter                    &  yes& our fiducial choice \\
pe-avsl-ac-lvg      &   pressure-entropy  &  strong limiter                       &  yes& lower order velocity gradients \\
pe-avwl       &   pressure-entropy  &  weak limiter             &  no &  \\
pe-avwl-ac    &   pressure-entropy  &  weak limiter             &  yes&  \\
pe-avsl-ac-erho &   pressure-entropy  &  strong limiter  &yes  & use $\rho^e$ in the dissipation terms\\
\hline\hline
\end{tabular}
\end{table*}

\subsection{Gresho vortex} 
The initial condition of the Gresho vortex test
\citep{1990IJNMF..11..621G,2010ARA&A..48..391S} consists of a
two-dimensional differentially rotating vortex. The velocity and
pressure profiles are as follows:  
\begin{eqnarray}
\begin{aligned}
	v_{\phi}(R) = 
	\begin{cases}
		5R     &\text{for } 0   \leq R < 0.2 , \\
		2 - 5R &\text{for } 0.2 \leq R < 0.4 , \\
		0      &\text{for } R   \geq 0.4 ;
	\end{cases} \\
\end{aligned}
\end{eqnarray}
\begin{eqnarray}\label{eq:greshoP}
\begin{aligned}
	P(R) = 	
	\begin{cases}
		5 + 12.5 R^2                     &\text{for } 0   \leq R < 0.2 , \\
		9 + 12.5 R^2 \\
		~ -20R + 4 \ln{(5R)}  &\text{for } 0.2 \leq R < 0.4 , \\
		3 + 4\ln{2}                      &\text{for } R   \geq
                0.4, 
	\end{cases}
\end{aligned}
\end{eqnarray}
where $R$ is the radius. The density is constant $\rho = 1$
everywhere. We set up a slab in a close-packing lattice with an
equivalent one-dimensional resolution $N$. The pressure profile is
devised to balance the centrifugal force, which makes the system
time-independent. 

\begin{figure}
	\centering
	\includegraphics[width=3.2in]{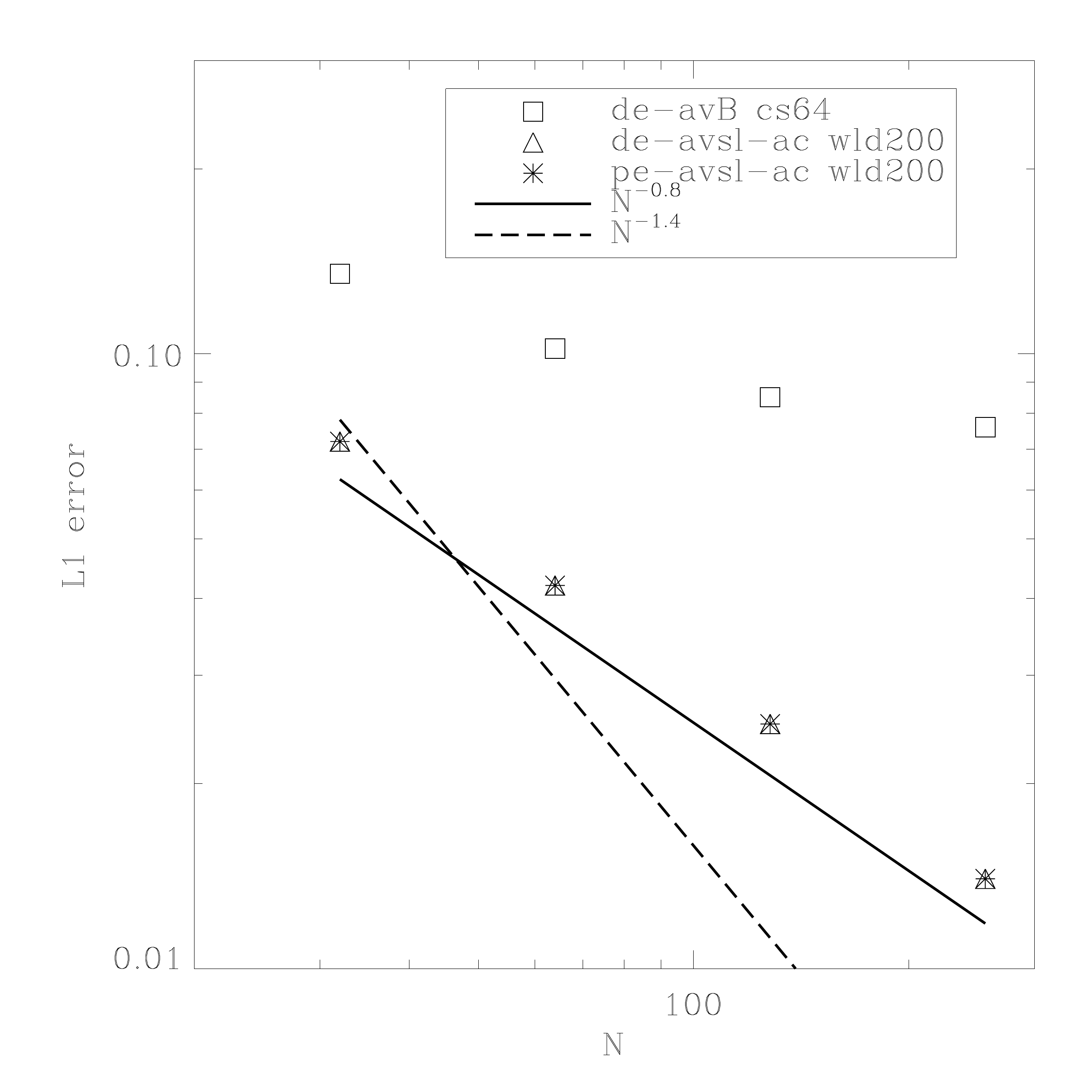}	
	\caption{Convergence rate (L1 error vs. particle number) for
          the standard Gresho test at t = 1. The DE formulation with the
          standard cubic spline and 64 neighbors (cs64, open squares)
          has the worst convergence properties. The use of the
          Wendland $C^4$ kernel with 200 neighbors (wld200) improves
          convergence significantly (L1 $\propto N^{-0.8}$). The PE
          formulation (asterisks) and the DE formulation (triangles)
          give almost identical results. The dashed line (L1 $\propto
          N^{-1.4}$) is the convergence rate found with the moving mesh code AREPO 
          \citep{2010MNRAS.401..791S}.} 
	\label{fig:gresho_conv_rate} 
\end{figure}

Fig. \ref{fig:gresho_conv_rate} shows the convergence rate (the L1
error) \footnote{We follow \citet{2010ARA&A..48..391S} and define the
  L1 error as the arithmetic average of the difference between the
  bin-averaged particle velocity and the analytic solution. The bin
  size we use is 0.01.} 
of different SPH schemes. Due to the use of the Wendland $C^4$ kernel the  
convergence rate ($\propto N^{-0.8}$) is significantly improved (see
\citet{2012MNRAS.422.3037R,2012MNRAS.425.1068D} for a more detailed
analysis) compared to the traditional SPH using cubic spline (see 
e.g. \citealp{2013MNRAS.428.1968K}) with only 64 neighbors, though 
still lower than for moving mesh methods (e.g. $\propto N^{-1.4}$
found in \citealp{2010MNRAS.401..791S}). \footnote{ One way to achieve a convergence rate $\propto N^{-1.4}$ for SPH, as shown in \citet{2012MNRAS.422.3037R}, is to factor out the E0 error. However, such a scheme violates momentum conservation and performs poorly at strong shocks, making it less favorable in most of applications.}
The DE and PE formulations
give almost exactly the same results. This is not surprising as both
density and pressure are continuous in this test and the two
formulations are equally accurate.

\begin{figure}
	\centering
	\includegraphics[trim = 5mm 20mm 25mm 30mm, clip,width=3.3in]{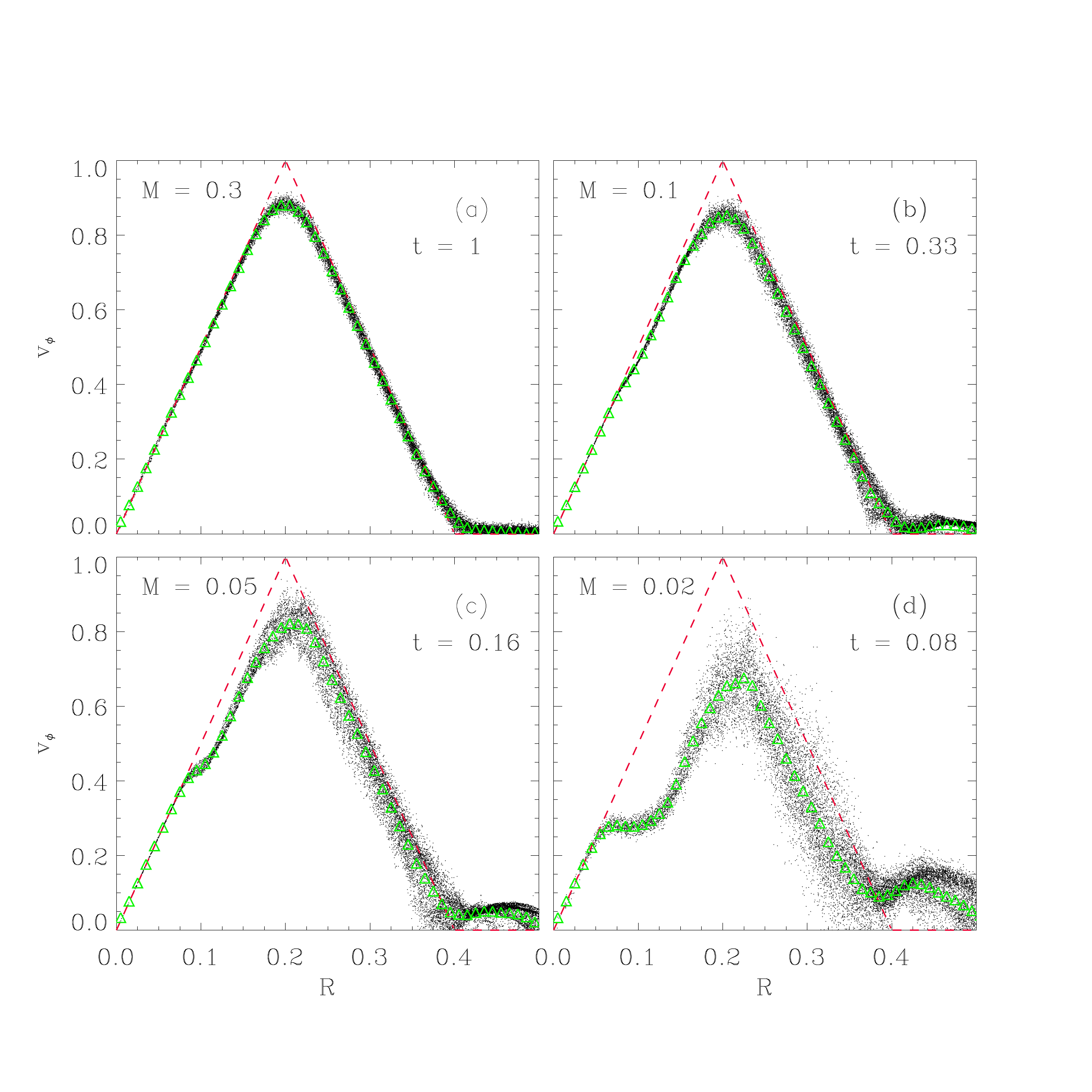}	
	\caption{The velocity profile in the Gresho test with Mach
          number $M$ = 0.3 (a), 0.1 (b), 0.05 (c), 0.02 (d), using our
          fiducial SPH implementation ('pe-avsl-ac'). The 1D
          resolution is $N$ = 256. The time is proportional to the
          sound-crossing time. Only randomly drawn one per cent of the
          particles are shown. A lower limit of the AV coefficient
          $\alpha_{\rm min}$ = 0.1 is used. The analytic solution is
          shown by the dashed red lines. The green triangles are the
          radial-binned mean values of $v_{\phi}$. As the Mach number
          decreases, the velocity profiles start to deviate from the
          analytic solution the scatter increases.} 
	\label{fig:greshoLowMach_amin01} 
\end{figure}

Recently, \citet{2013PhD} presented a more general form of the
standard Gresho problem which allows for changing the maximum Mach
number of the system to study the behavior of grid-based methods in
the low Mach number regime. While the density and velocity profiles
remain unchanged, the pressure is modified as
\begin{eqnarray}
\begin{aligned}
	P(R) = 	
	\begin{cases}
		P_0 + 12.5 R^2                     &\text{for } 0   \leq R < 0.2 , \\
		P_0 + 12.5 R^2 + 4  \\
		~ -20R + 4 \ln{(5R)}  &\text{for } 0.2 \leq R < 0.4 , \\
		P_0 - 2 + 4\ln{2}                      &\text{for } R
                \geq 0.4, 
	\end{cases}
\end{aligned}
\end{eqnarray}
where $P_0 = (\gamma M^2)^{-1}$, and $M$ is the maximum Mach number of the system.
Note that this is merely a constant shift with respect to Equation
(\ref{eq:greshoP}). Therefore, the pressure gradient force does not
change and still balances the centrifugal force. For $M = \sqrt{3/25}
\approx 0.3$ it recovers the original Gresho problem. We now can
investigate the performance of our fiducial SPH scheme (PE formulation
with AC) in the low Mach number
regime. Fig. \ref{fig:greshoLowMach_amin01} shows the results with
different Mach number $M$ with 1D resolution $N$ = 256.
The times are in code units and are proportional to the sound-crossing time for a given Mach number.
The $M$ = 0.3 case (corresponds to the standard Gresho test) in panel
(a) is recovered nicely, though the systematically biased values at
the discontinuities ($R$ = 0.2 and $R$ = 0.4) are still unavoidable. 
However, as the Mach number decreases, the scatter becomes more severe
and the radial-binned mean values start to deviate from the analytic
expectation. This is especially notable in the region of rigid-body
rotation ($R < 0.2$). At lower Mach numbers the viscous force become
stronger due to a higher sound speed. Therefore, even though the AV
coefficient is maintained at a minimum value, the viscous effect is
still conspicuous, leading to a spurious transportation of angular
momentum. In addition, the fluctuations due to the E0 error are also
more severe for higher sound speeds, which explains the scatter. 

\begin{figure}
	\centering
	\includegraphics[trim = 5mm 20mm 25mm 30mm, clip, width=3.3in]{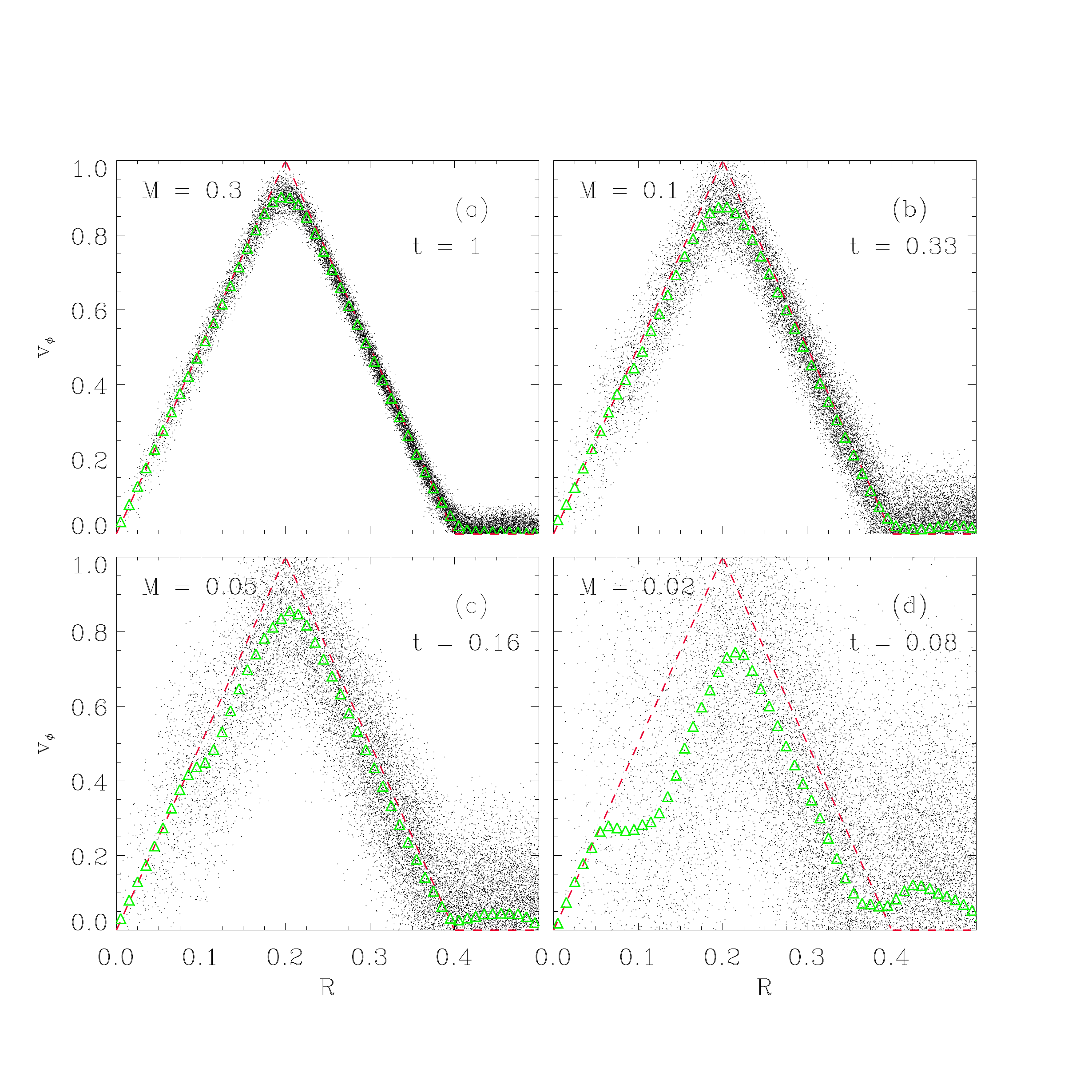}	
	\caption{Same as Fig. \ref{fig:greshoLowMach_amin01} but
          with no lower limit to the AV coefficient. The mean values
          (green triangles) are improved only moderately while the scatter 
          increases significantly.} 
	\label{fig:greshoLowMach_no_amin} 
\end{figure}

One plausible 'solution' is to remove the lower limit of the AV
coefficient to reduce the viscous effects completely. In
Fig. \ref{fig:greshoLowMach_no_amin} we repeat the same tests but
set $\alpha_{\rm min}$ = 0. The viscosity is then only triggered by
our AV switch. Unfortunately, the radially binned mean values are
improved only modestly (slightly higher peak values) compared to
Fig. \ref{fig:greshoLowMach_amin01} while the scatter increases
significantly. The particle fluctuation caused by the E0 error is much
more severe without the minimum viscosity. Such fluctuations in turn
triggers AV which then still leads to the deviation of the mean
values. The effort of reducing the spurious transportation of angular
momentum is hence in vain. We therefore use a minimum viscosity as our
default choice. 

In summary, due to the E0 error, it remains challenging to model
highly subsonic shear flows ($M \leq 0.05$) even with our favored SPH
scheme. This is in agreement with the results of
\citet{2012MNRAS.423.2558B} who points out that the traditional SPH
scheme yields problematic results in subsonic turbulence. This seems
to be an intrinsic issue of SPH as the E0 error can only be reduced by
using larger number of neighboring particles if the conservation
properties are kept \citep{2012JCoPh.231..759P, 2012MNRAS.425.1068D}. 

\subsection{Sod shock tube}\label{sec:sod}
\begin{figure}
	\centering
	\includegraphics[trim = 13mm 0mm 0mm 0mm, clip, width=3.3in]{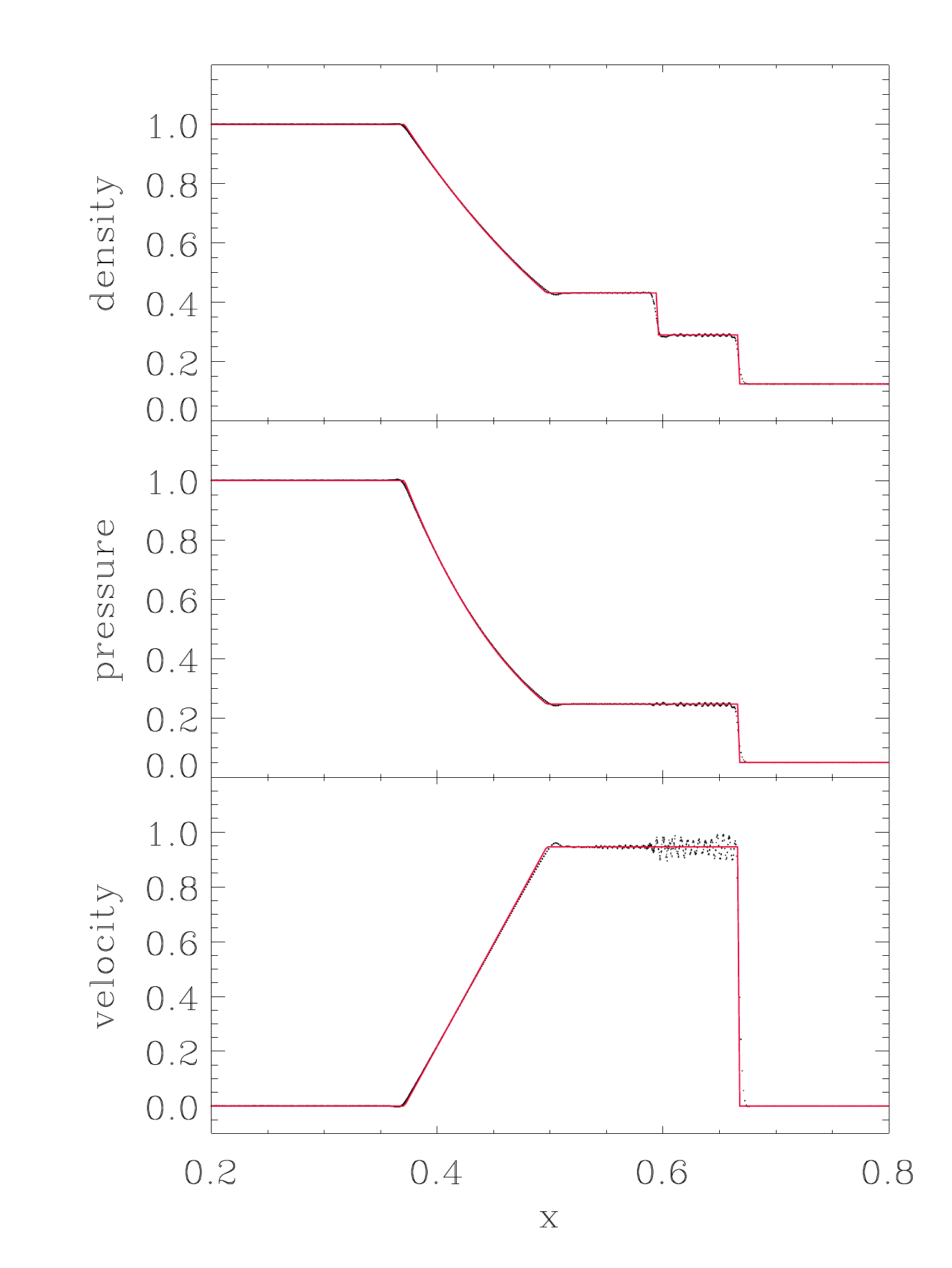}
	\caption{Shock tube test at t = 0.1, using our fiducial SPH
          ('pe-avsl-ac'). The density and pressure (top and middle
          panels) are in good agreement with the analytic prediction
          (red curves), while the velocity profile (bottom panel)
          shows weak oscillations in the post-shock region. Overall
          our fiducial scheme passes this test very well.} 
	\label{fig:sod} 
\end{figure}

In this section we present results from a shock tube test
\citep{1978JCoPh..27....1S} to assess the accuracy of the schemes for
weak shocks. This standard test consists of two fluids that are initially
stationary with moderate discontinuities in both density and pressure at
the interface, producing three characteristic waves, namely the shock,
the contact discontinuity, and the rarefaction wave, respectively. We
set the initial density $\rho_l = 1$ and pressure $P_l = 1$ on the
left-hand side, and $\rho_r = 0.125$ and $P_r = 0.05$ on the
right-hand side, which makes the shock slightly supersonic (with a
Mach number M $\approx$ 1.5). We set up a 3D tube with the total
length $L$ = 1 in a close-packing lattice with effectively 600
particles along the tube. The polytropic index is $\gamma = 5/3$. We
do not initialize the AV coefficient to its maximum at the interface. 
Fig. \ref{fig:sod} shows the result of the shock tube test at t =
0.1 with our fiducial SPH scheme ('pe-avsl-ac'). Both the density and
pressure profiles are in good agreement with the analytic predictions.  
The locations of the three characteristic waves are correctly modeled. 
There are however some oscillations in the velocity profile in the
post-shock region which probably indicates too little viscosity. This
error should be acceptable as long as the average value still agrees
with the analytic solution. 

\begin{figure}
	\centering
	\includegraphics[trim = 10mm 0mm 5mm 0mm, clip, width=3.in]{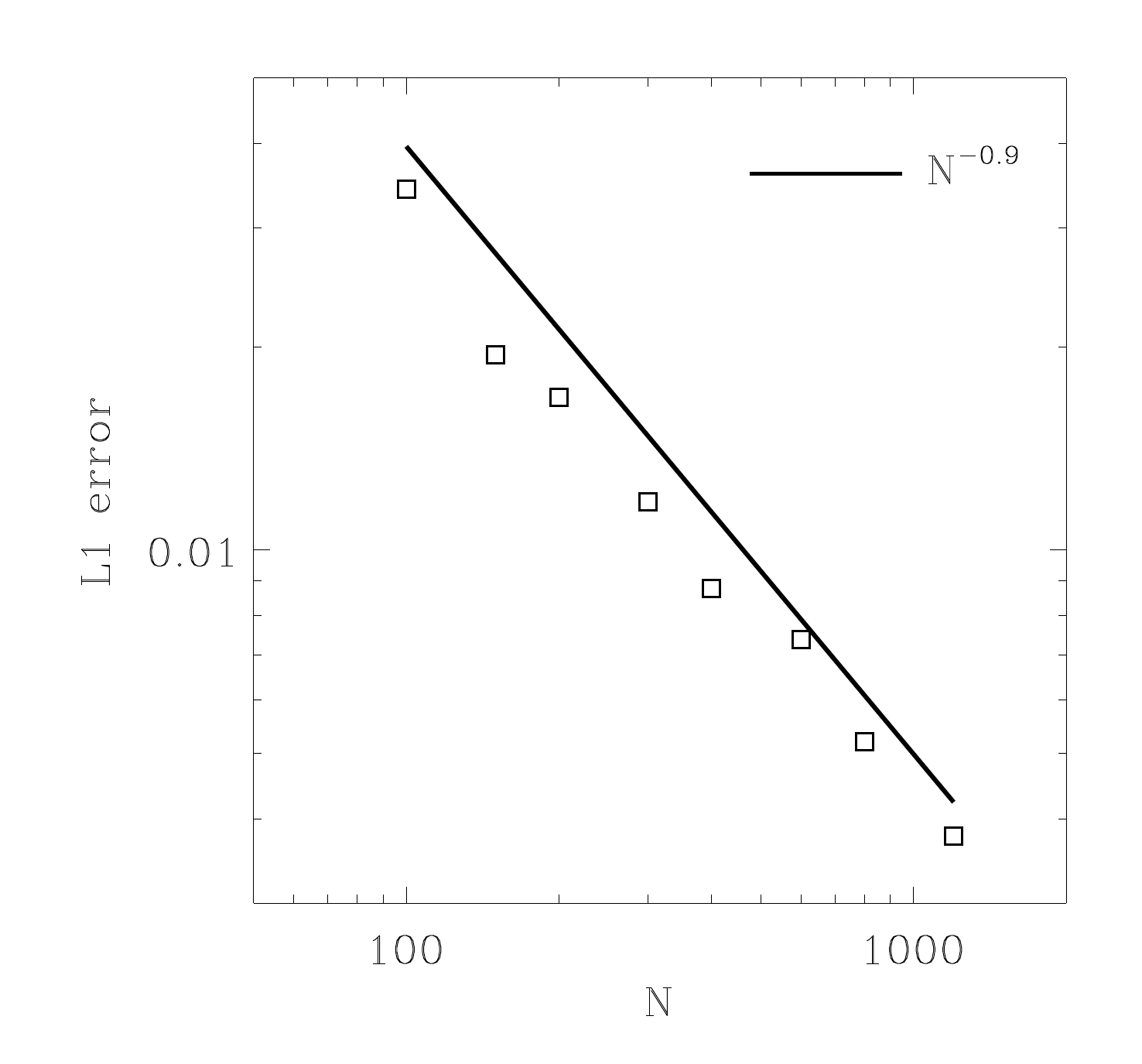}
	\caption{The convergence rate of the shock tube test at t = 0.1, using our fiducial SPH ('pe-avsl-ac'). The convergence rate is close to $L1 \propto N^{-0.9}$, close to the optimal rate of $N^{-1}$ (see also \citealp{2010ARA&A..48..391S}).}
	\label{fig:sod_conv} 
\end{figure}

We further investigate the convergence rate of the velocity profile in Fig. \ref{fig:sod_conv}.
Following \citet{2012MNRAS.422.3037R}, we define the L1 error as 
\begin{equation}
	L1 = \frac{1}{N_b} \sum_{i}^{N_b} |{\bar v_{x,i}} - {\bar v_x(x_i)}|
\end{equation}
where $N_b$ is the number of bins, ${\bar v_{x,i}}$ is the mean value of the particle velocity in the $x$-direction within bin $i$, and ${\bar v_x(x_i)}$ is the mean analytic solution within bin $i$. We set a binsize of 0.005 and exclude the bins without particles. This binsize is small enough to capture the post-shock oscillations even in our highest resolution. We find the convergence rate close to $L1 \propto N^{-0.9}$, which is in agreement with \citet{2010ARA&A..48..391S, 2012MNRAS.422.3037R}. As discussed in \citet{2010ARA&A..48..391S}, the accuracy of the scheme is limited to first order due to the error at the discontinuity.

\subsection{Sedov explosion}\label{sec:sedov}
The Sedov explosion \citep{1959sdmm.book.....S} is an ideal test to
evaluate the shock capturing capability of the code under extreme
entropy contrasts. We set up a three-dimensional cube with $128^3$
particles in a glass-like distribution. The initial density is
constant at $\rho = 1.24 \times 10^7 \rm{M}_{\odot} kpc^{-3}$ ($\rm
n_H \approx 0.5 cm^{-3}$) and the temperature is $T \approx 6.5 ~\rm{K}$.  
The central 64 particles are then injected
uniformly with a thermal energy of $E = 6.78 
\times 10^{53} \rm ergs$, which results in a top-hat profile with a
huge entropy contrast of $3\times 10^6$. This corresponds to a Mach
number $M \approx 1000$. We set the polytropic index to $\gamma = 5/3$
which represents the non-radiative phase of a Sedov blast wave
(e.g. \citealp{1988RvMP...60....1O}). This is a standard setup for
testing a very strong shock (e.g. \citealp{2013MNRAS.428.2840H}).

\begin{figure}
	\centering
	\includegraphics[trim = 15mm 10mm 10mm 20mm, clip, width=3.6in]{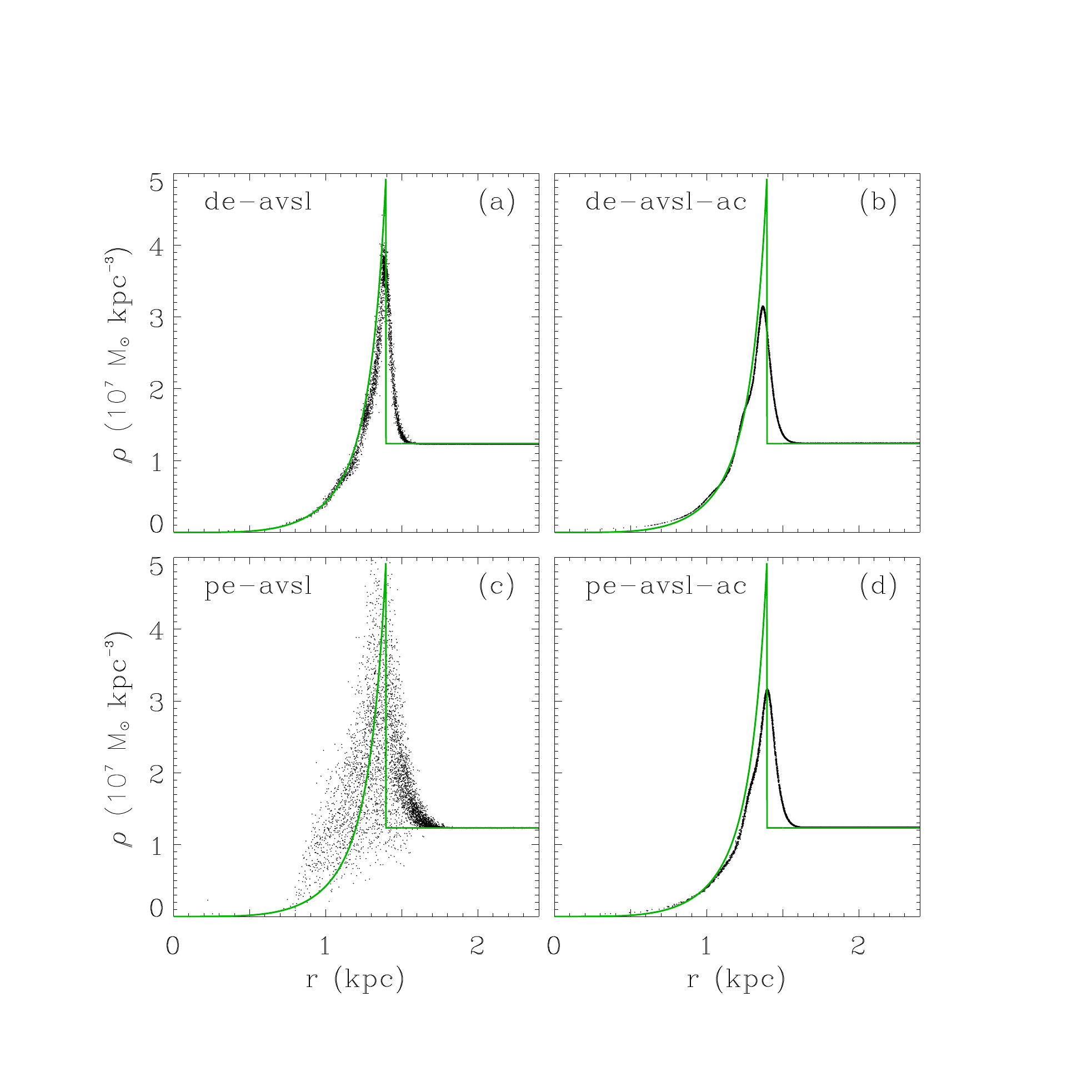}
	\caption{Density profile for a Sedov explosion after t = 30
          Myr (showing only two per cent of randomly drawn particles).  
	{\it Panel (a):} DE formulation without AC. {\it Panel (b):}
        DE formulation with AC. {\it Panel (c):} PE formulation without AC.
	{\it Panel (d):} PE formulation with AC.
	The green curves are the analytic solution. Without AC, the
        profile matches the analytic solution reasonably well (despite
        some scatter) using the DE formulation. On the other hand,
        using the PE formulation without AC the profile is very much
        dominated by noise. Including AC improves the results
        dramatically for both formulations (b and d).}
	\label{fig:sedov_density} 
\end{figure}

Fig. \ref{fig:sedov_density} shows the density profiles at t = 30 Myr
using the DE and PE formulation, respectively. We plot two per cent of
randomly drawn particles instead of radial binned values to indicate
the scatter. For the DE formulation, the density profile agrees with
the analytic solution quite well, though without AC (a) the scatter is
higher. This is due to the sharp entropy discontinuity in the initial
conditions. Once we include AC (b), the scatter is reduced significantly. 
This is in agreement with results presented in \citet{2012MNRAS.422.3037R}.

For the PE formulation the situation becomes worse since the pressure
estimate is entropy-weighted. Without AC
(Fig. \ref{fig:sedov_density}, panel c), particles penetrate the shock
front anisotropically and eventually the noise dominates.  This can in
principle be improved by increasing the maximum value of the AV
coefficient. 
\textbf{
Indeed, \citet{2013ApJ...768...44S} and \citet{2013MNRAS.428.2840H} reported reasonably well results of a similar test without AC but using a much higher AV coefficient.
}
However, we find that including the AC (d) already
significantly reduces the scatter.  More importantly, for a reason
that will be explained in Section \ref{sec:holes}, we try to avoid too
much AV.  The AC softens entropy jumps at the shock front and makes
the entropy more continuous within the kernel.  The resulting density
profile and its scatter is as good as using the DE formulation with
AC. We therefore consider AC a necessary ingredient in the PE
formulation when strong shocks are involved. This is especially true
for the disk simulations in Section \ref{sec:disk} since our cooling
implementation allows gas to cool down to a few hundred Kelvin instead
of $10^4$ K and the entropy contrast is usually much larger.

\begin{figure}
	\centering
	\includegraphics[trim = 10mm 10mm 10mm 20mm, clip, width=3.6in]{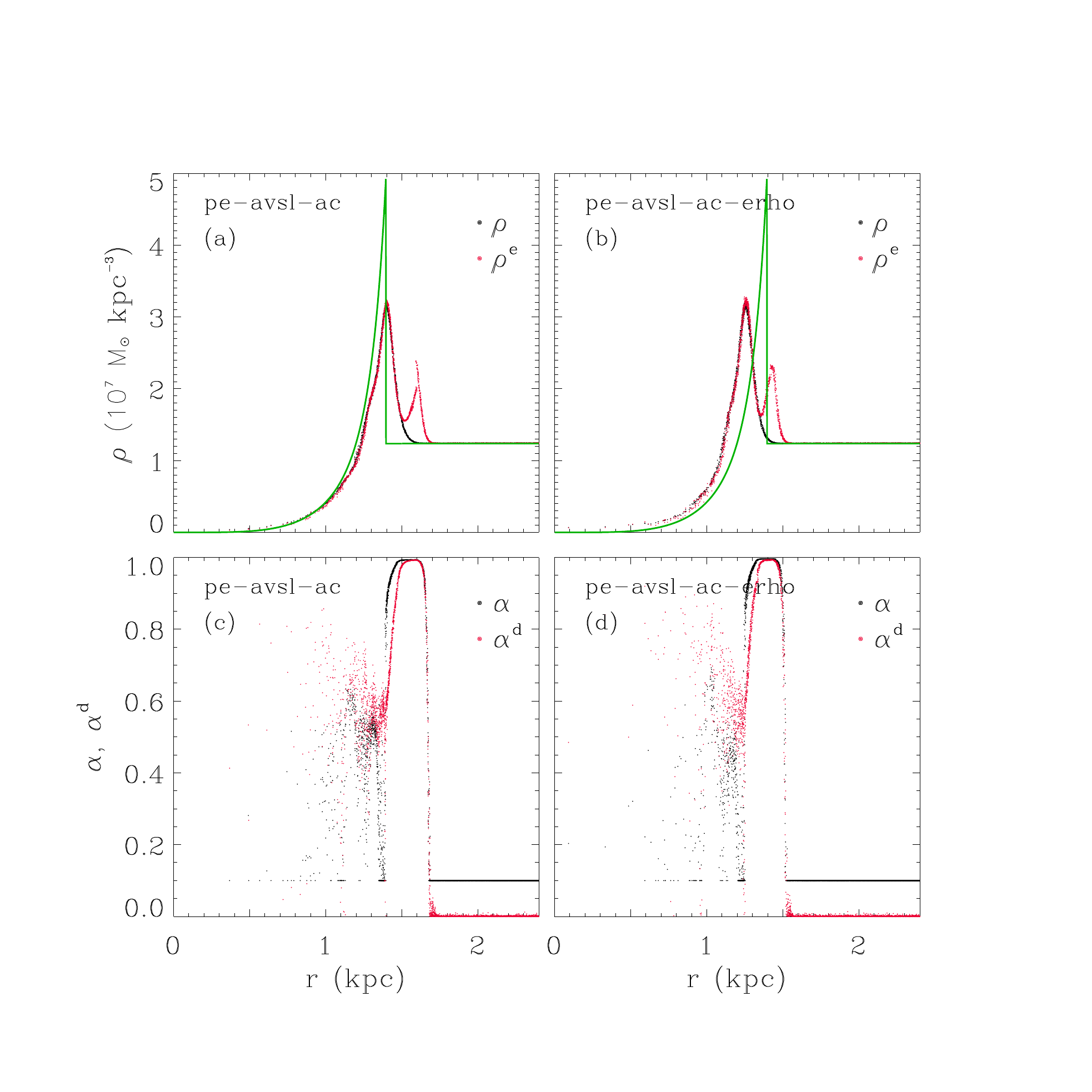}
	\caption{Density profiles and the distributions of the AV and
          AC coefficients at t = 30 Myr, using different definitions of
          density in the dissipation terms (showing only two per cent of
          randomly drawn particles). {\it Panel (a):} density profile using
          mass-weighting in the dissipation terms. Black dots and red
          dots are the mass-weighted and entropy-weighted densities,
          respectively. The green curve is the analytic solution. The
          peak position matches the analytic solution. A pre-shock
          density bump can been seen in the entropy-weighted density. 
	{\it Panel (b):} same as (a) but using the entropy-weighted density
        in the dissipation terms. The peak falls behind the analytic
        solution. {\it Panel (c):} distribution of the AV (black dots)
        and AC (red dots) coefficients corresponding to (a).	 
	{\it Panel (d):} distribution of AV and AC coefficients
        corresponding to model (b).}
	\label{fig:ewrho} 
\end{figure}

As discussed in Section \ref{sec:AV}, the density can be defined in
two different ways for the PE formulation: as a mass-weighted density
or a entropy-weighted density.  In Fig. \ref{fig:ewrho}, panel (a), we 
show the two density estimates for the same simulation.  At the shock
($\sim$ 1.2 kpc) the two densities agree well with each
other. However, the entropy-weighted density has a "bump" ahead of the
shock front. This bump can be understood as follows: in the Sedov
test the entropy is discontinuous but the pressure is smoothed.  
Therefore, the entropy-weighted density $\rho^e_i = (\widehat{P}_i /
A_i)^{1/\gamma}$ would be over/under-estimated in the pre/post-shock
regions, biasing the result.  Since the entropy jump is several orders
of magnitude, the bias in the pre-shock region is much more severe and
manifests itself as a bump. This is very similar to the "pressure
blip" problem in the DE formulation.  Both stem from the fact that one
variable (density or pressure) is being smoothed while the other
(entropy) remains sharply discontinuous.

In Fig. \ref{fig:ewrho}, panel (b), we further compare to a simulation
using the entropy weighted density in the dissipation terms (AV and AC).  
Here the shock front falls notably behind. This can be understood as
the result of the spurious pre-shock bump.  In the pre-shock region
the artificial viscosity is switched on, converting kinetic energy
into thermal energy.  However, the over-estimated density causes the
entropy to be under-estimated in Equation (\ref{eq:ent_AV}). As such,
the shock loses part of its driving force as if some energy has been
lost or radiated away, and thus propagates slower.  As shown in
Fig. \ref{fig:ewrho}, panels (c) and (d), the AV and AC 
are both efficiently switched on right before the shock arrives. There
are a few oscillations in the post-shock region as a result of the
post-shock ringing and the limiter in Equation (\ref{eq:limiter}).  

\begin{figure}
	\centering
	\includegraphics[trim = 10mm 10mm 10mm 20mm, clip, width=3.6in]{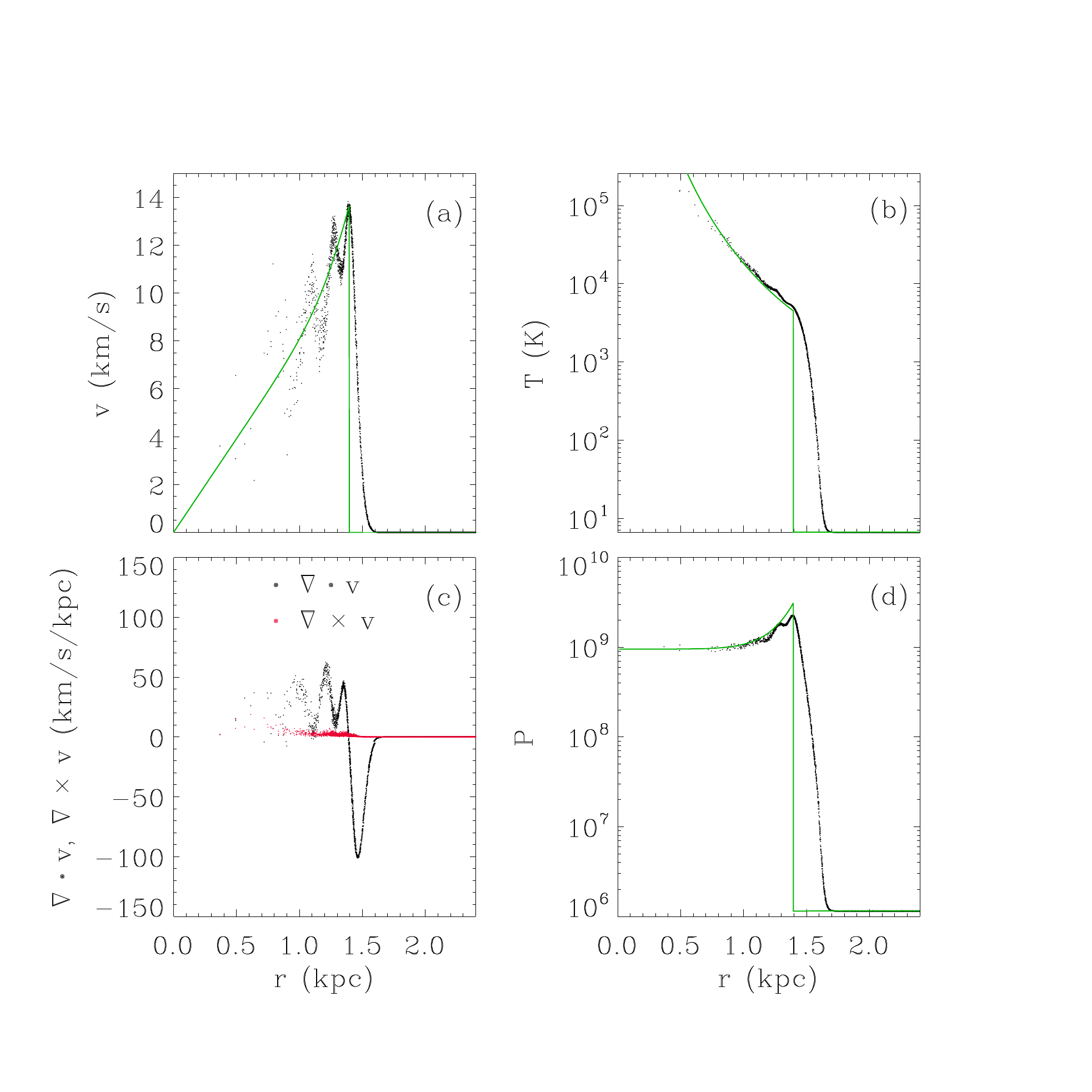}
	\caption{Radial profile of velocity (a), temperature (b),
          divergence and curl of velocity (c), and pressure (d). All
          panels are generated with our fiducial SPH scheme
          ('pe-avsl-ac'). Only two per cent of randomly drawn
          particles are shown. The results are in good agreement with
          the analytic solution (green curves).}  
	\label{fig:sedov_vTP} 
\end{figure}

Fig. \ref{fig:sedov_vTP} shows the radial profile of velocity,
temperature, and pressure for our fiducial model as well as the
analytic solutions. Here, the post-shock ringing can be seen most
clearly in the velocity profile and is unfortunately difficult to
avoid in our implementation.  The radial profile of the velocity
divergence and curl shows that the former always dominates over the
latter and the particle order in the post-shock region is kept
reasonably well.

\subsection{Keplerian ring}
\label{subsec:Keplerian}
In this section we discuss results for a Keplerian ring test similar
to \citet{2009MNRAS.395.2373C,2010MNRAS.408..669C}. We set up a two
dimensional ring with a Gaussian surface density profile 
\begin{equation}
	\Sigma(r) = \frac{1}{m} \exp{[-\frac{(r-r_0)^2}{2\sigma^2}]},
\end{equation}
where $m$ is the total mass of the ring, $r_0$ is the radius of the
peak surface density, and $\sigma$ is the width of the ring.  We set
$r_0 = 10$ and $\sigma = 1.25$. The ring is subject to the gravity of
a point mass $M \gg m$ located at $r = 0$. The self-gravity of the
ring is neglected and we set $GM = 1000$, where $G$ is the
gravitational constant. The initial condition is set up with 9987
particles using the method presented in \citet{2009MNRAS.395.2373C}
which generates a lattice-like particle distribution with concentric
rings.  

The rotation velocity of the ring follows a Keplerian velocity profile
$v_{\phi}(r) = \sqrt{GM/r}$ and the rotation period at $r_0$ is
therefore $T = 2\pi$. The sound speed of the ring, $c_s = 0.01 \ll
v_{\phi}$, ensures that inviscid hydrodynamical processes only set in
after several periods of rotation as discussed in
\citet{2010MNRAS.408..669C}. The AV can be falsely triggered by a
poor estimate of velocity gradients and then a viscous instability
develops quickly, breaking the ring structure. In this test we do not
use a minimum value of AV since even a small viscosity would lead to
the instability. Because of the low sound speed, a small perturbation
in velocity could be supersonic leading to shocks heating up the
ring. This mostly starts from the inner edge of the ring where
differential rotation is most prominent. In the absence of AV, the
system is in equilibrium and should remain stable over several periods
of rotation. Since the system is isothermal, the AC would not be
triggered and can be neglected for this test.

\begin{figure}
	\centering
	\includegraphics[trim =10mm 10mm 0mm 28mm, clip, width=3.8in]{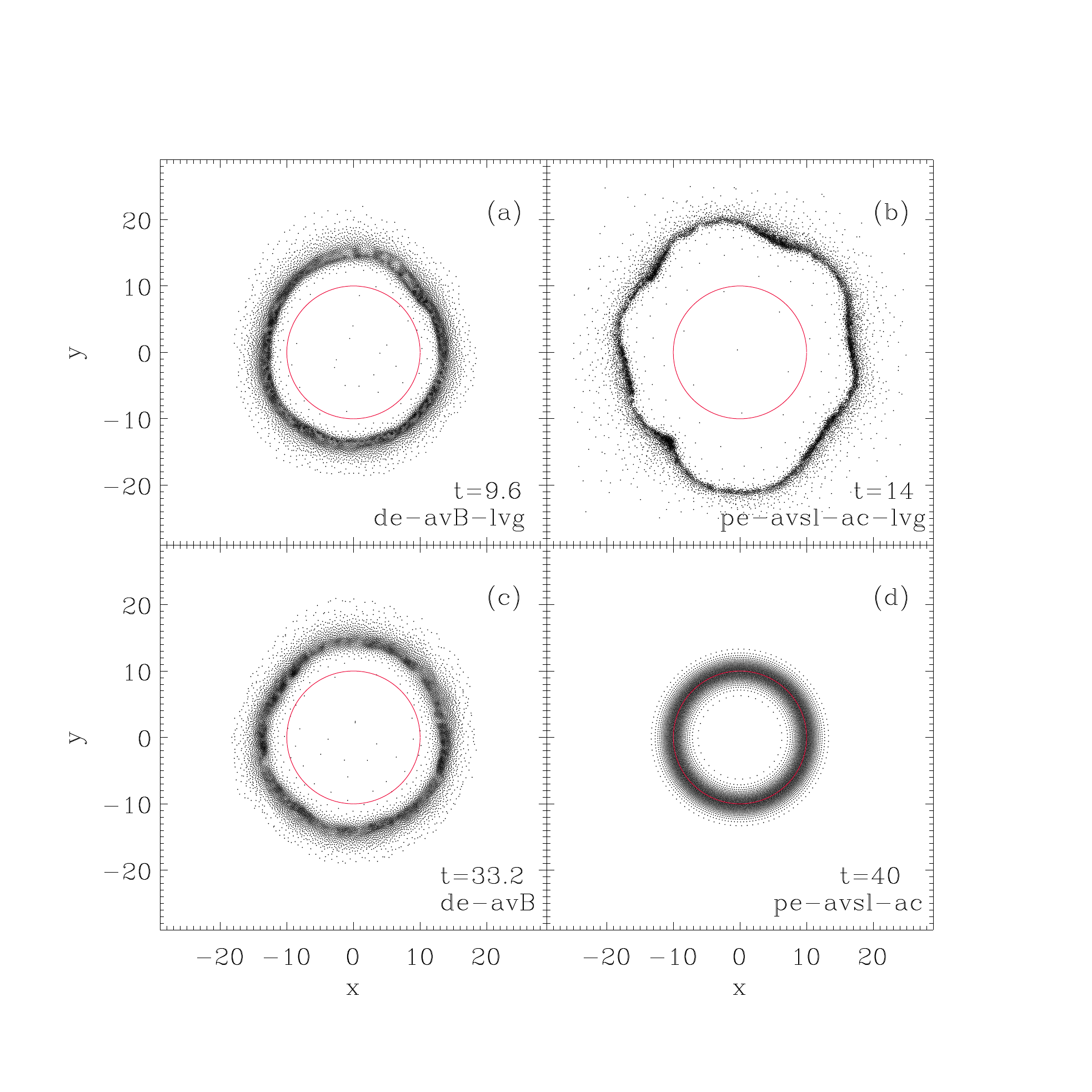}
	\caption{Two dimensional ring with a radial Gaussian density
          profile in Keplerian rotation at the time (in code
          units) when the ring structure breaks up. No lower limit of
          AV is used in this test. One rotation takes
          $T=2\pi$. {\it Panel (a):} DE-formulation with Balsara
          switch, using lower order velocity gradients. {\it Panel
            (b):} PE-formulation with AC, using lower order velocity
          gradients. {\it Panel (c):} DE-formulation with Balsara
          switch, using higher order velocity gradients. {\it Panel
            (d):} PE-formulation with AC using higher order velocity
          gradients. Using higher order velocity gradients prevents
          falsely triggered AV \citep{2010MNRAS.408..669C} and
          therefore the ring remains stable for more than five
          rotation periods. The red solid rings indicate the location of 
          the initial peak density.
          	}
	\label{fig:KD} 
\end{figure}

Fig. \ref{fig:KD} shows the results with different SPH schemes at
the time (in code units) after which the ring structure breaks up.
The standard Balsara switch (panel a) is too viscous and the
instability sets in very quickly after one and a half  rotation
periods ($T = 2 \pi$). Using our fiducial scheme but without the
higher order estimate of velocity gradients (panel b) only delays the
instability for less than one rotation period.  Adopting the higher order
velocity gradient estimator \citep{2010MNRAS.408..669C} greatly
improves the situation. Even with the Balsara switch that is usually
considered to be too viscous, the system is able to evolve for about
five periods before the instability sets in (panel b). With our
fiducial scheme the AV is further reduced, and the system remains
stable during the whole simulation time (t=40), very similar to the
implementation presented in \citet{2010MNRAS.408..669C}.

\subsection{Hydrostatic equilibrium test}
\begin{figure}
	\centering
	\includegraphics[width=3in]{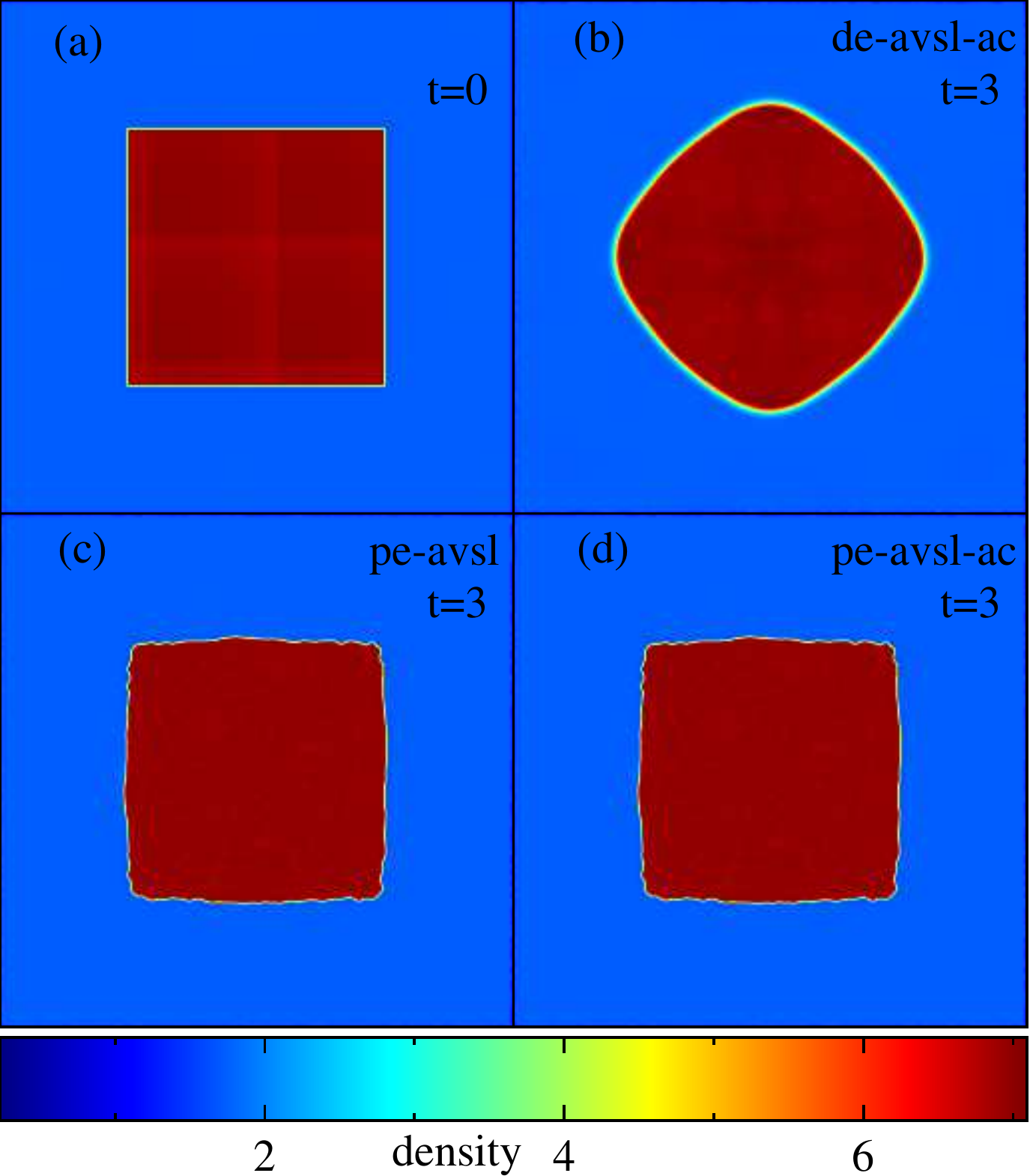}
	\caption{Density distribution for a square in hydrostatic
          equilibrium. {\it Panel (a):} initial condition. {\it Panel
            (b):} DE-formulation with AC at t = 3. 
	{\it Panel (c):} PE-formulation without AC at t = 3.
	{\it Panel (d):} PE-formulation with AC at t = 3.
	The DE-formulation leads to deformation of the square even
        with AC. With PE-formulation the surface tension is absent
        right from the beginning and therefore the square retains its
        original shape.} 
	\label{fig:box} 
\end{figure}

In this section we present results from a hydrostatic equilibrium test
similar to \citet{2013ApJ...768...44S,2013MNRAS.428.2840H} to examine
spurious surface tension at contact discontinuities. We set up a two
dimensional square $0 \leq x < 1$ and $0 \leq y < 1$ with periodic 
boundary conditions, filled up with a background fluid of uniform
density $\rho = 7/4$.  A slightly denser fluid of uniform density
$\rho = 7$ is embedded in the central region where $0.25 \leq x <
0.75$ and $0.25 \leq y < 0.75$.  We use $256\times 256\times 3$
particles in the background region and $512\times 512$ in the central
region, both in a cubic lattice configuration. All particles have the
same mass. The pressure $P = 3.75$ is constant in both fluids. Fig.
\ref{fig:box} shows the results for different SPH schemes after
following the evolution of the system up to $t=3$ (in code units). Initially the central fluid forms a perfect
square. It has been shown that the square evolves into a circle due to
spurious surface tension in standard SPH implementations
\citep{2013ApJ...768...44S,2013MNRAS.428.2840H}. Somehow surprisingly,
it also deforms significantly with the DE-scheme and AC. This is
probably due to the inertia  exerted by the spurious surface tension
in the initial conditions. Using the PE-formulation with or without AC
the square retains its original shape (as demonstrated in 
\citealp{2013ApJ...768...44S,2013MNRAS.428.2840H}), although there are
some small fluctuations at the boundaries because of the E0-error
force.  In the PE-formulation, the spurious surface tension is absent
right from the beginning.

\subsection{Kelvin-Helmholtz instability}
\begin{center}
\begin{figure*}
\includegraphics[trim = 0mm 25mm 0mm 0mm, clip, width=8in]{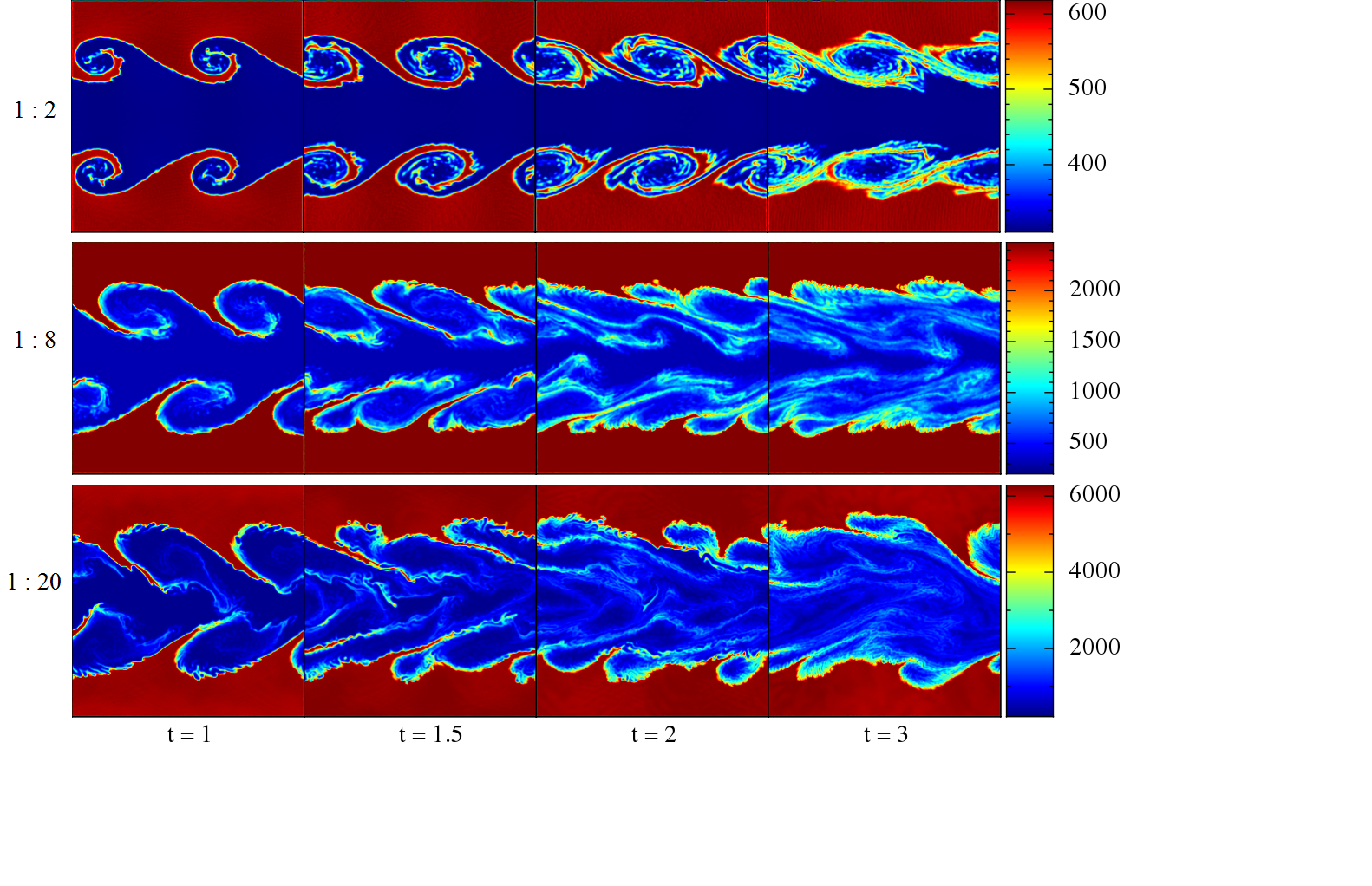}
\caption{The density distribution (in units of ${\rm
    M}_{\odot}/{\rm kpc}^3$) at z = 8 kpc for the KH
  instability test at t = 1, 1.5, 2, 3 $\tau_{\rm KH}$ (from left
  to right) with a density contrast of 1:2 (top row), 1:8 (middle
  row), and 1:20 (bottom row). The instability develops in a few
  characteristic time-scales thanks to the PE formulation. 
  }
\label{fig:kh3in1}
\end{figure*}
\end{center}

The Kelvin-Helmholtz (KH) instability develops at contact
discontinuities of shear flows and is an important mechanism for the
onset of fluid turbulence.  The inaccuracy of standard SPH in this
test and the solutions to it have been well studied in the literature
(e.g. \citealp{2007MNRAS.380..963A, 2008JCoPh.22710040P,
  2010MNRAS.407.1933J, 2012A&A...546A..45V, 2013MNRAS.428.1968K,
  2012MNRAS.422.3037R, 2013ApJ...768...44S, 2013MNRAS.428.2840H}).  
Here we simply show the results with our fiducial scheme. The initial
condition we set up is identical to \citet{2012MNRAS.422.3037R}. The
computational domain is a periodic slab of $256\times 256\times 16$
kpc. We use 774144, 2359296, and 5304748 equal-mass particles in a
cubic lattice for density contrast of 2, 8, and 20, respectively. 
The fluid is divided into three (the top, middle and bottom) parts,
which are in pressure equilibrium. The middle layer has a density 
$\rho_1 = 313 ~{\rm M}_{\odot}/{\rm kpc}^3$ and temperature $T_1 =
10^7$ K. The top and bottom layer have a density $\rho_2 = \chi
\rho_1$ and temperature $T_2 = T_1 / \chi$ K where $\chi$ is the
density contrast. The shear flow is set up such that the central fluid 
travels at $v_x$ = 40 km/s while the upper and lower fluid travel at
$v_x$ = -40 km/s. A sinusoidal velocity perturbation in y direction
of $\delta v_y$ = 4 km/s with the wavelength $\lambda$ = 128 kpc is
applied to trigger the instability. The instability is expected to
develop within a few Kelvin-Helmholtz time scales  
\begin{equation}
	\tau_{\rm{KH}} \equiv \frac{(1 + \chi)\lambda}{(\chi)^{1/2} v},
\end{equation}
where $v$ = 80 km/s is the relative shear velocity at the interface. 
Fig. \ref{fig:kh3in1} shows the density contours at t = 1, 1.5, 2
and 3 $\tau_{\rm KH}$ (from left to right) for a density contrast of
$\chi = 2$ (top row), $\chi = 8$ (middle row), and $\chi = 20$ (bottom
row). The KH instability develops in qualitative agreement with
\citet{2012MNRAS.422.3037R} and \citet{2013MNRAS.428.2840H}. 
Note that the PE formulation itself is already capable of resolving
the KH instability and the AC here is suppressed by the limiter in
Equation (\ref{eq:limiter}) to avoid unnecessary dissipation.
However, for higher density contrasts (20:1) there are some
small-scale fluctuations at the boundaries due to the E0 error force
and the system becomes more diffusive.

\subsection{Blob test}
\begin{figure}
	\centering
	\includegraphics[width=3.5in]{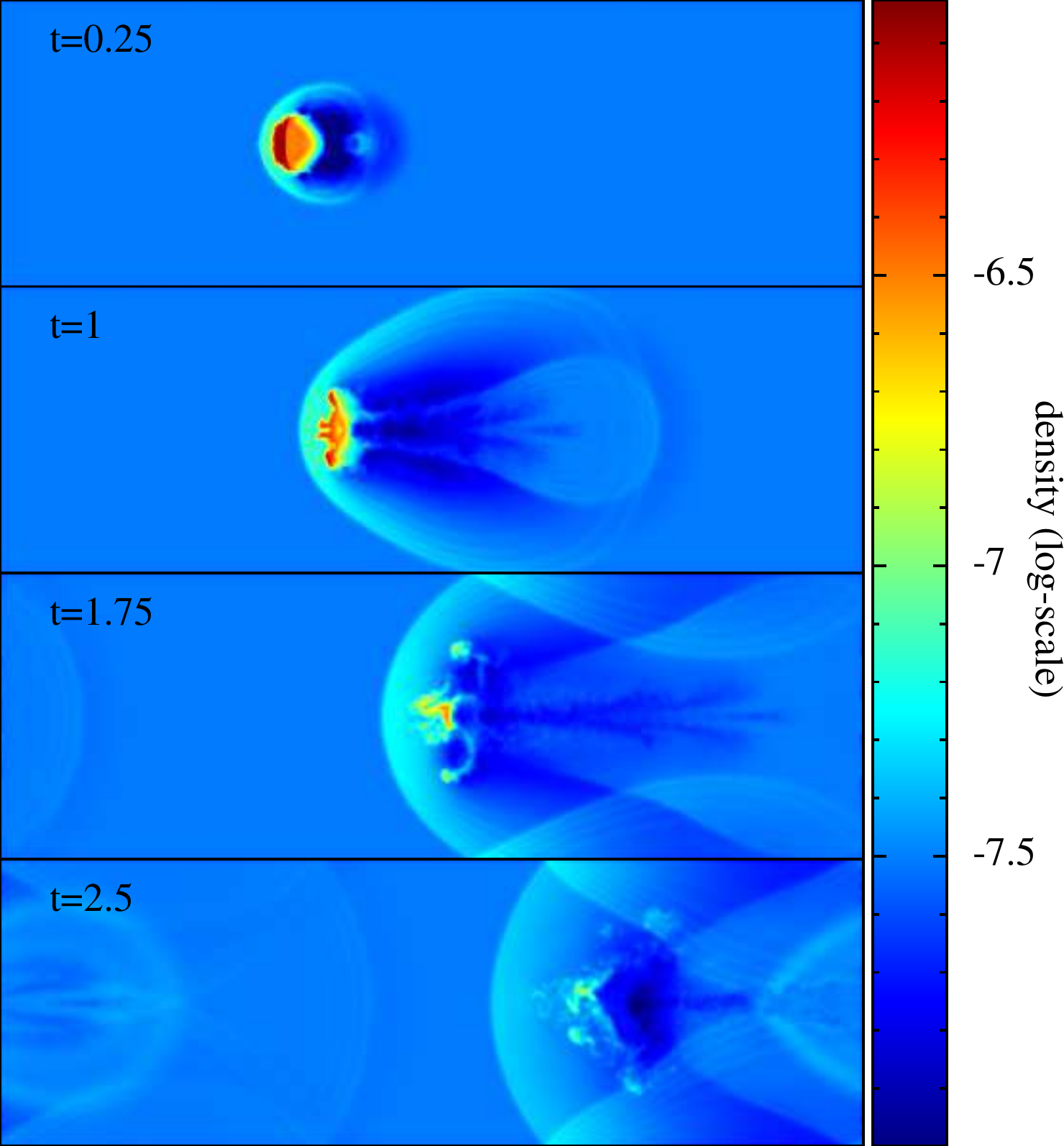}
	\caption{Density distribution (in log-scale) of a central
          slice at t = 0.25, 1, 1.75, 2.5 $\tau_{\rm KH}$ (from top to
          bottom) for the blob test, using our fiducial SPH scheme
          ('pe-avsl-ac'). The cloud dissolves within a few $\tau_{\rm
            KH}$ and the results are in qualitative agreement with
          grid-based methods as well as other improved SPH
          implementations.}  
	\label{fig:blob}
\end{figure}

The blob test \citep{2007MNRAS.380..963A} is a complicated problem
which involves several important physical processes.  In this test, a
spherical cloud of radius $R_{\rm cl}$ travels with Mach number $M =
2.7$ in an ambient medium that is 10 times hotter and 10 times less
dense.  We use the same initial conditions \footnote{Available at
  http://www.astrosim.net/code/doku.php} as presented in
\citet{2010MNRAS.405.1513R}.  The computational domain is a tube with
a size of $10\times 10\times 30$ in units of the cloud radius, with
periodic boundary conditions. The particle number is 9641651 and all
particles have the same mass. Spherical harmonics are used to
initialize the perturbations of the cloud in order to trigger the
instabilities. Although no analytic solution exists for such problem,
the qualitative behavior can be useful to assess the capability of the
code to mix fluids. In Fig. \ref{fig:blob} we show the density
contours (in log-scale) at t = 0.25, 1, 1.75 and 2.5 $\tau_{\rm KH}$
using our fiducial SPH scheme.  The cloud dissolves within a few
$\tau_{\rm KH}$ and our results are in qualitative agreement with
grid-based  methods \citep{2007MNRAS.380..963A} as well as other
improved SPH implementations \citep{2010MNRAS.405.1513R,
  2012MNRAS.422.3037R,2013MNRAS.428.2840H}. 

\section{Isolated disk galaxy}\label{sec:disk}
In this section we present simulations of a more realistic application
to a real astrophysical problem: the dynamical evolution of an
isolated disk galaxy. These simulations model additional physical
processes that are discussed below.  

We use the method outlined in \citet{2005MNRAS.361..776S} to set up
the initial conditions (see also \citealp{2011MNRAS.415.3750M}). The
galaxy consists of a stellar and gaseous disk component with a total
mass of $3.9\times 10^{10} \rm{M_{\odot}}$ and a gas fraction of
0.2. The radial scale-length of the stellar and gaseous exponential
profile is 2.5 kpc. The stellar disk has a scale height of 0.6
kpc. The vertical structure of the gaseous disk is determined
assuming hydrostatic equilibrium. In addition, the galaxy has a
stellar bulge  ($9.7\times 10^{9} \rm{M_{\odot}}$) following a
Hernquist \citep{1990ApJ...356..359H} profile with a scale length of
0.3 kpc.  
The disk galaxy is embedded in a dark matter halo with a 
virial radius $r_{\rm vir}$ = 160 kpc and a mass 
$M_{\rm halo} = 1.3\times 10^{12} \rm{M_{\odot}}$.
The dark matter halo follows also a Hernquist profile with an NFW-equivalent \citep{1997ApJ...490..493N} concentration parameter $c$ = 9 such that the scale length $r_s = r_{\rm vir}/c$.
The particle numbers for the different components are $N_{\rm halo} = 3\times 10^6$ for the dark matter halo, $N_{\rm disk} = 3\times 10^6$ for the stellar disk, $N_{\rm bulge} = 7.5\times 10^5$ for the stellar bulge and $N_{\rm gas} = 6\times 10^5$ for the gaseous disk.
The softening lengths are 68 pc for the dark matter halo and 13 pc for the gas, disk and bulge components.
We also set up an initial radial metallicity 
gradient of -0.04 dex/kpc and $Z=Z_{\odot}$ at 8 kpc from the galactic
center.  All details of the model are given in Table \ref{table:IC}.

\begin{table}
\caption{Parameters for the isolated disk galaxy}
\label{table:IC}
\begin{tabular}{| c | c | c | c |}    
\hline\hline
$M_{\rm halo}$  &  $1.3\times10^{12}$  & $\rm{M_{\odot}}$  &  halo mass  \\
$N_{\rm halo}$  &  $3\times 10^6$     &     &  halo particle number \\
$v_{\rm cir}$   &  160        & km/s      &  halo circular velocity\\
$c$             &  9          &             &  halo concentration    \\
$\lambda$       &  0.035      &             &  spin parameter        \\
\hline
$M_{\rm disk}$  &  $3.1\times10^{10}$ &$\rm{M_{\odot}}$  &  stellar disk mass\\
$N_{\rm disk}$  &  $3\times 10^6$     &     &  disk particle number\\
$r_{\rm disk}$  &  2.5                  &kpc&  disk scale length\\
$h_{\rm disk}$  &  0.6                &kpc&  disk scale height \\
\hline
$M_{\rm gas}$  &  $7.8\times10^9$ &$\rm{M_{\odot}}$  &  gas mass\\
$N_{\rm gas}$  &  $6\times 10^5$     &     &  gas particle number\\
$r_{\rm gas}$  &  2.5                  & kpc &  gas scale length\\
$d(\log Z)/dr$ & -0.04  & dex/kpc & metallicity gradient\\
\hline
$M_{\rm bulge}$  &  $9.7\times10^9$ &$\rm{M_{\odot}}$  &  bulge mass bulge\\
$N_{\rm bulge}$  &  $7.5\times 10^5$     &     &  bulge particle number\\
$r_{\rm bulge}$  &  0.3                  &kpc&  bulge scale length\\
\hline\hline
\end{tabular}
\end{table}

\subsection{Cooling and star formation}
\label{sec:extraPhys}
We adopt the implementation of metal enrichment and cooling presented
in \citet{2013MNRAS.434.3142A}, which is partly based on
\citet{2006MNRAS.371.1125S}. We trace 11 individual elements H, He, C,
N, O, Ne, Mg, Si, S, Ca and Fe for all gas and star particles. These
elements are produced by chemical enrichment of SNII, SNIa and AGB
stars and then advected with the particles.  For gas particles, we
include metal diffusion (see \citealp{2013MNRAS.434.3142A} for details)
to account for turbulent mixing.  
The radiative cooling rate of the gas is computed on an
element-by-element basis, assuming optically thin gas in ionization
equilibrium under the UV/X-ray background as in
\citet{2009MNRAS.393...99W}.  The minimum temperature is set to $10^2$ K.

We adopt a standard estimate for the local star formation rate:
\begin{equation}
	\frac{{\rm d}\rho_{\star}}{\rm dt} = \epsilon \frac{\rho_{\rm{gas}}}{\rm t_{dyn}},
\end{equation}
where ${\rm t_{dyn}}$ is the local dynamical time $(4\pi G \rho_{\rm
  gas})^{-0.5}$, $\epsilon$ is the star formation efficiency, 
$\rho_{\star}$ and $\rho_{\rm gas}$ are the volumetric density of
stellar and gas component respectively.  We take $\epsilon$ = 0.04 as
our default choice. The star formation threshold is set to a number
density of $n^{\rm th}_{\rm gas} \geq 1 ~\rm{cm}^{-3}$ and we require
the temperature to be $T \leq 10^4$ K. Gas particles that are either
too hot or too dilute will not participate in the star formation 
process.  

For the stellar feedback we only consider mass, momentum and energy
input from SN explosions. We assume that the SN events (about  $\sim$ 
3 Myr after a stellar particle has formed) transfer mass, radial
momentum, and thermal energy to the  nearest 10 gas particles in a
free-expansion approximation \citep{1988RvMP...60....1O}. The mass of
the supernova ejecta (typically $\sim 19$ per cent of the formed
stellar population) is directly added to its neighbors and distributed
in a kernel-weighted fashion. Each individual supernova explosion
injects $10^{51}$ ergs into the ISM, which corresponds to an ejecta
velocity $v_{\rm e} \sim$  3000 km/s.  The momentum is transferred to
the neighbors similar to an inelastic collision and the "post-shock"
gas velocity is   
\begin{equation}
	{\bf v}_{\rm gas}^{\prime} = \frac{ m_{\rm gas} {\bf v}_{\rm
            gas} + \Delta m {\bf v}_e }{ m_{\rm gas} + \Delta m}. 
\end{equation}
The remaining energy is transformed into thermal energy and added to
the affected gas particles 
\begin{equation}
	\Delta U = \frac{1}{2} \frac{m_{\rm gas}\Delta m}{ m_{\rm gas}
          + \Delta m} |{\bf v}_e - {\bf v}_{\rm gas}|^2. 
\end{equation}
This feedback implementation conserves mass, momentum and energy
simultaneously.  The injected momentum is small (as we do not assume a
Sedov approximation) but it will not be radiated away easily, so the
feedback can be more efficient, especially in the dense clumps of
gas. As we will show below this feedback implementation naturally
drives outflows from the galactic disk.

\subsection{Evolution of the gaseous disk}\label{sec:holes}
It has been shown that the method for setting up
initial conditions presented in \citet{2005MNRAS.361..776S} produce
stable models for disk galaxies, also in the presence of gas
(see e.g. \citealp{2010MNRAS.403.1009M,2012MNRAS.423.2045M}). We refer
to these papers for all necessary details. In this section we only
focus on the impact of 
a specific SPH implementation on the gas phase properties while
keeping the star formation algorithm fixed. For this comparison we
model the dynamical evolution for $\sim$ 1 Gyr and investigate the
morphology of the gas disk and the rates of star formation,
outflow from the disk.

As mentioned in Section \ref{sec:AV}, we adopt a 'strong limiter' in our
fiducial AV scheme as opposed to a 'weak limiter' first explored by
\citet{2010MNRAS.408..669C}. 
The motivation is that the AV scheme with a weak limiter seems to be
too viscous in our isolated disk simulations.  In this section we
study the impact of these two different limiters on the evolution of
the gaseous disk, particularly the distribution of the gas density.

\subsubsection{Density-entropy formulation}\label{sec:disk_de}
\begin{figure}
	\centering
	\includegraphics[width=3in]{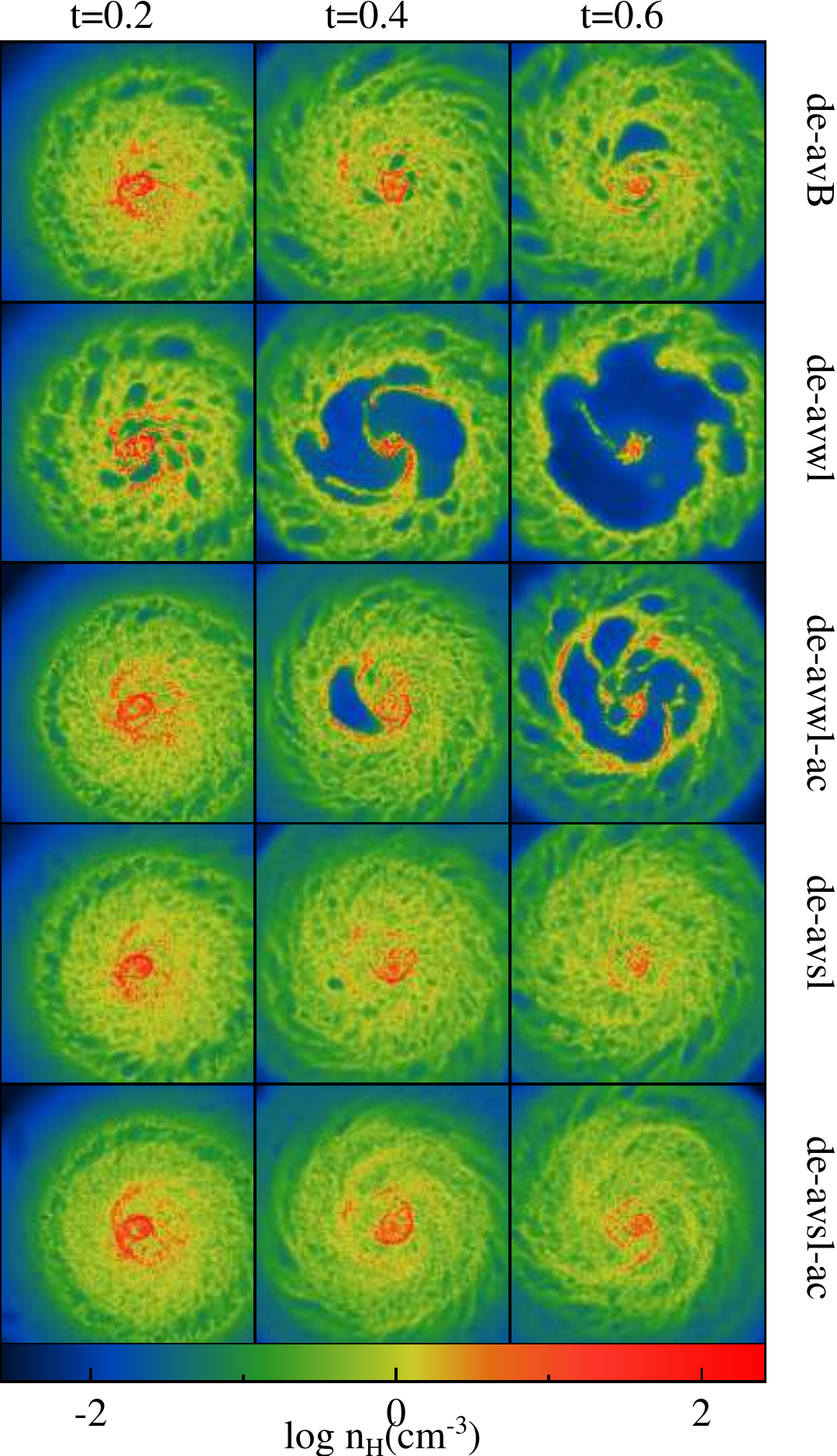}
	\caption{The projected density of gas (face-on) in the
          isolated disk simulation with different SPH schemes (from
          top to bottom) at t = 200 (left column), 400 (middle
          column), and 600 (right column) Myrs respectively. The SPH
          schemes all use the DE formulation but with different
          dissipation schemes. {\it Top row:} Constant AV with a
          Balsara switch creates holes that gradually merge but are
          sheared away. {\it Second row:} With weak AV limiter the
          holes merge and the instability evacuates most of the gas in 
          the disk. Including AC mitigates the situation (third
          row). {\it Two bottom rows:} The AV switch with a strong limiter
          makes the disk gas less viscous and the instability does not
          develop, both with and without AC . 
    }
	\label{fig:disk_de}
\end{figure}

In Fig. \ref{fig:disk_de} we show the evolution of the gas density
(face-on) using the density-entropy (DE) formulation of SPH with different
dissipation schemes. The standard constant viscosity with a Balsara
switch is shown in the top row. We find that some small "holes" form
in the central part of the disk and gradually grow to kpc size and
then start to be sheared away. They seem to be related to the stellar
feedback as the SN explosions create bubbles in the surrounding ISM
(though on a much smaller scale). However, similar holes also form at
large radii, where star formation is unimportant. 
The question is whether
the formation and growth of these bubbles through merging is physical
and whether it is caused by feedback or is an artifact of too much
viscosity due to the numerical implementation, similar to the
Keplerian ring (see Section \ref{subsec:Keplerian}).   
We therefore test the AV switch with a weak limiter, which does not 
lead to instabilities in the  Keplerian ring. 
However, the situation becomes even more problematic on a shorter time 
scale (second row of Fig. \ref{fig:disk_de}). After 600 Myrs almost
the entire gas is evacuated from the disk which is  clearly too severe
to be realistic.  The situation is improved if we  include AC (third
row). However, it only delays the instability for a few Myrs.

With our fiducial AV with a strong limiter these holes disappear both
without AC (fourth row) as well as with AC (bottom row). If we include
the AC the disk becomes smoother and clear spiral arms are formed.   
As discussed in Section \ref{sec:AV} the main feature of the 'strong
limiter' implementation is that the viscosity coefficient is always 
smaller than the limiter $\xi_i$ (times the constant maximum value
$\alpha_{\rm max}$ which we set to be 1).  

\begin{figure}
	\centering
	\includegraphics[width=3.in]{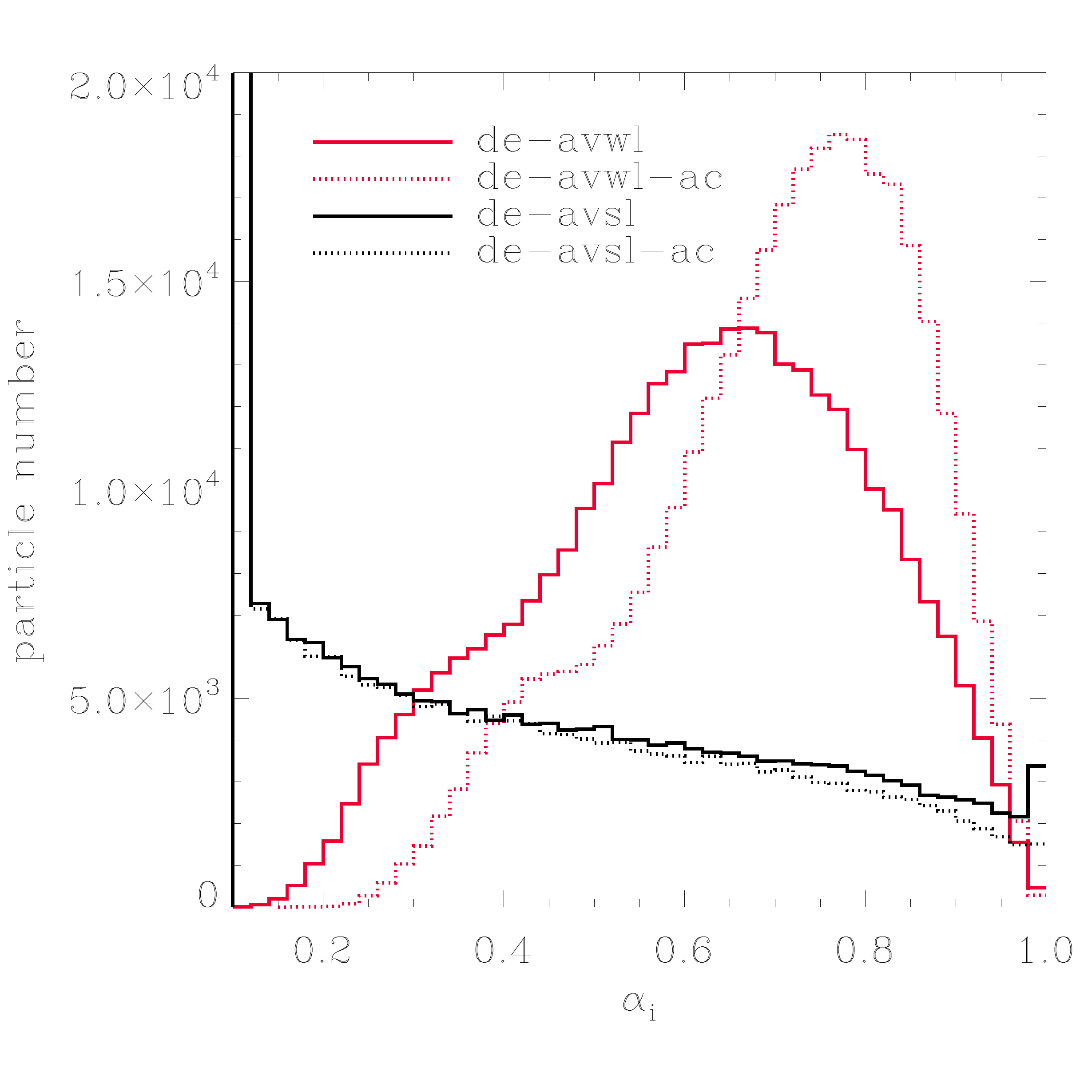}
	\caption{Histogram of the AV coefficient in the isolated disk
          simulation at t = 300 Myr using the DE-formulation. The AV
          coefficient with a weak limiter (red solid and dotted lines)
          peaks at 0.6-0.8, while with a strong limiter (black solid
          and dotted lines) most particles retain the minimum value $\alpha_i=0.1$. 
	}
	\label{fig:hist_avis_de}
\end{figure}

We interpret the evolution of the 'avwl' models 
as a viscous instability, eventually triggered by SN-induced
blast waves.  In Fig. \ref{fig:hist_avis_de} we show the histograms of
the viscosity coefficients for all gas particles in the weak and
strong limiter case. In case of the weak limiter the disk is clearly
much more viscous. The viscosity coefficient is above the minimum
value of $0.1$ for almost all particles and peaks at around $\alpha_i
= 0.8$. With a strong limiter the majority of the particles retain their
minimum value $\alpha_i = 0.1$.  We therefore interpret the
instability as a result of having too much viscosity.  The physical
explanation is the following: in the low viscosity (strong limiter)
case, the small holes created by the SN explosions are quickly sheared
away before they have the chance to merge.  On the
other hand, if the viscosity is generally too high or decays too
slowly in the post-shock  regions, the holes become too viscous to be
sheared away by the differential rotation.  As such, they merge 
and trigger even more shocks, leading to the catastrophic
instability. The strong limiter efficiently suppresses viscosity in
the rotating disk and thus avoids such instability. 

\subsubsection{Pressure-entropy formulation}
\begin{figure}
	\centering
	\includegraphics[width=3in]{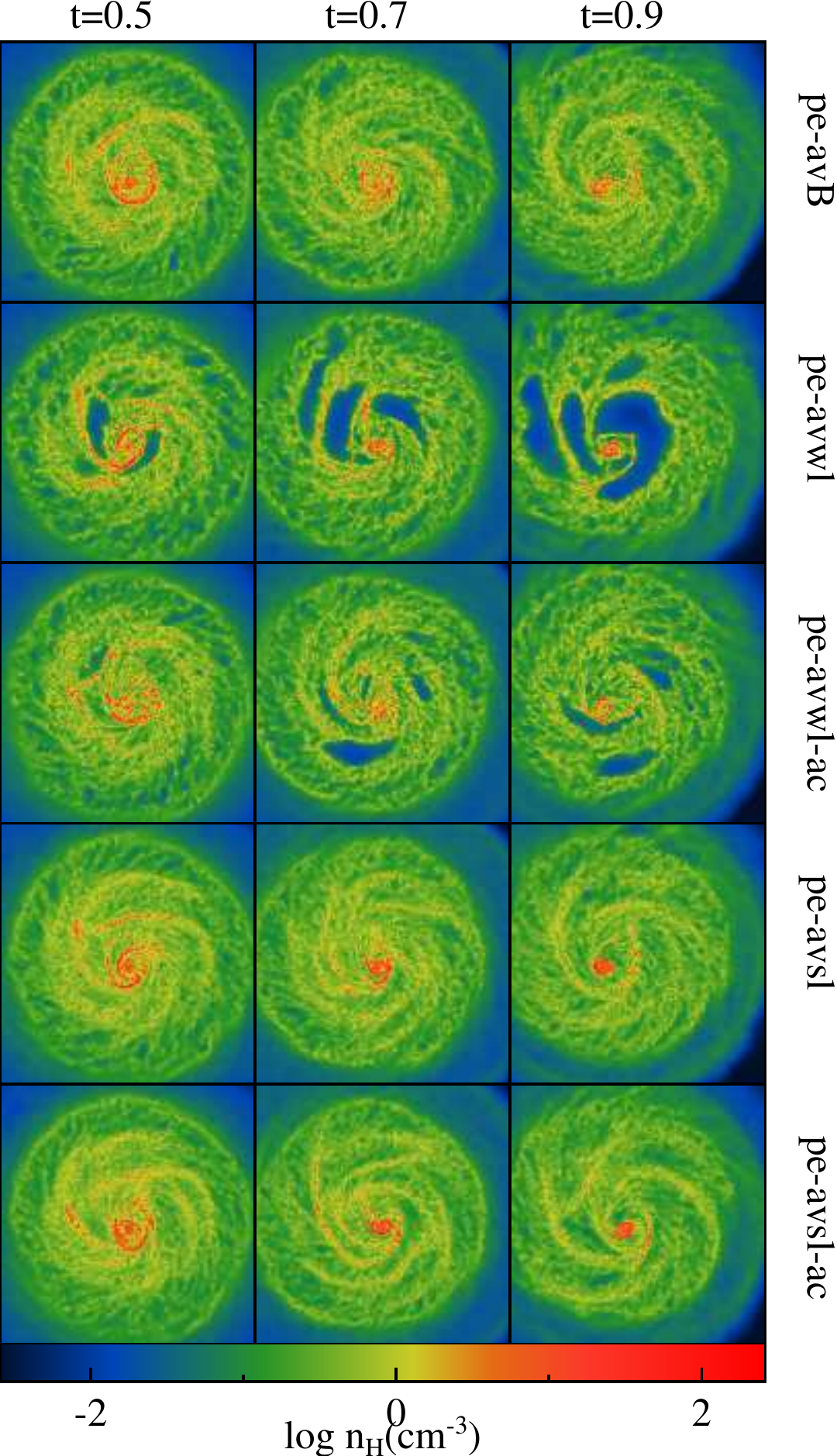}
	\caption{Same as Fig. \ref{fig:disk_de} but for the PE
          SPH scheme at t = 500 (left column), 700 (middle column),
          and 900 (right column) Myrs. The PE formulation significantly
          alleviates the instability. However, large holes still form
          using a weak limiter, as a result of the disk gas being too
          viscous.}  
	\label{fig:disk_pe}
\end{figure}

\begin{figure}
	\centering
	\includegraphics[width=3.in]{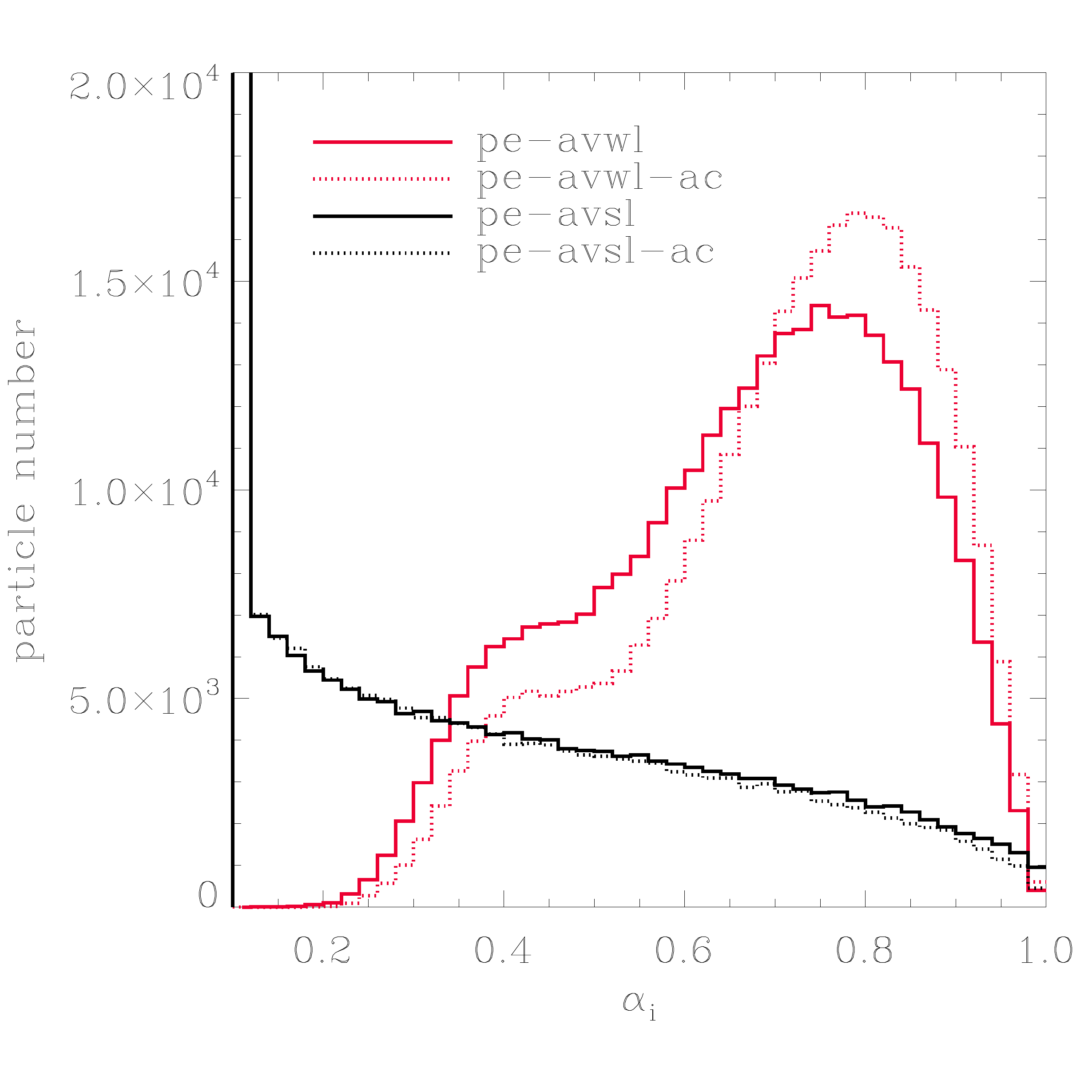}
	\caption{Same as Fig. \ref{fig:hist_avis_de} but using the PE
          scheme with no obvious differences from the DE scheme.}
	\label{fig:hist_avis_pe}
\end{figure}

We perform the same set of simulations as in Section \ref{sec:disk_de}
but now with the PE formulation.  Fig. \ref{fig:disk_pe} shows the
results at t = 500, 700, and 900 Myr.  The onset of the instability is 
significantly delayed with the PE formulation in all cases.  With a weak
AV limiter, the disk remains stable for a much longer time (until t =
500 Myr) than for the DE case.  However, it still becomes
unstable if we run the simulations up to 1 Gyr. For constant viscosity
(plus Balsara switch, top row of Fig. \ref{fig:disk_pe}) as well as
the AV switch with a strong limiter (two bottom rows of
Fig. \ref{fig:disk_pe}), the disk remains stable at all times and
distinct spiral arms are formed.  Fig. \ref{fig:hist_avis_pe} shows
the corresponding histogram of the viscosity coefficient for different AV
schemes which looks very similar to the DE implementation (Fig. \ref{fig:disk_de}).
This suggests that the artificial surface tension in the DE
formulation also plays a role for the onset and evolution of the
instability. In the DE formulation, the boundaries between hot and 
dilute post-shock bubbles and its surrounding cold and dense ISM are
difficult to break once being created.

We tentatively conclude that the instability is a consequence of
having too much viscosity in the disk, supported by spurious surface
tension. 
Including AC or using the PE-formulation both alleviate the situation.
However, the best "solution" is to avoid too much viscosity in the
first place. On the other hand, having too little viscosity is also
dangerous as it is required for proper shock modeling. The key is therefore
to introduce an appropriate amount of viscosity in the "right" places.  
Our AV scheme (with a strong limiter) delivered reassuring results for
the Sedov explosion test. However, in galactic disks where both
shocks and shear flows coexist, it is unclear how much viscosity
should be introduced. This is a general issue for most SPH schemes
that rely on AV for shock capturing. 
An exception might be 'Godunov SPH' (e.g. \citealp{2010MNRAS.403.1165C},
\citealp{2011MNRAS.417..136M}) which does not rely on artificial viscosity for shock capturing.

\subsection{Thermaldynamic properties}\label{sec:pd}
\begin{figure}
	\centering
	\includegraphics[trim = 5mm 20mm 5mm 25mm, clip, width=3.6in]{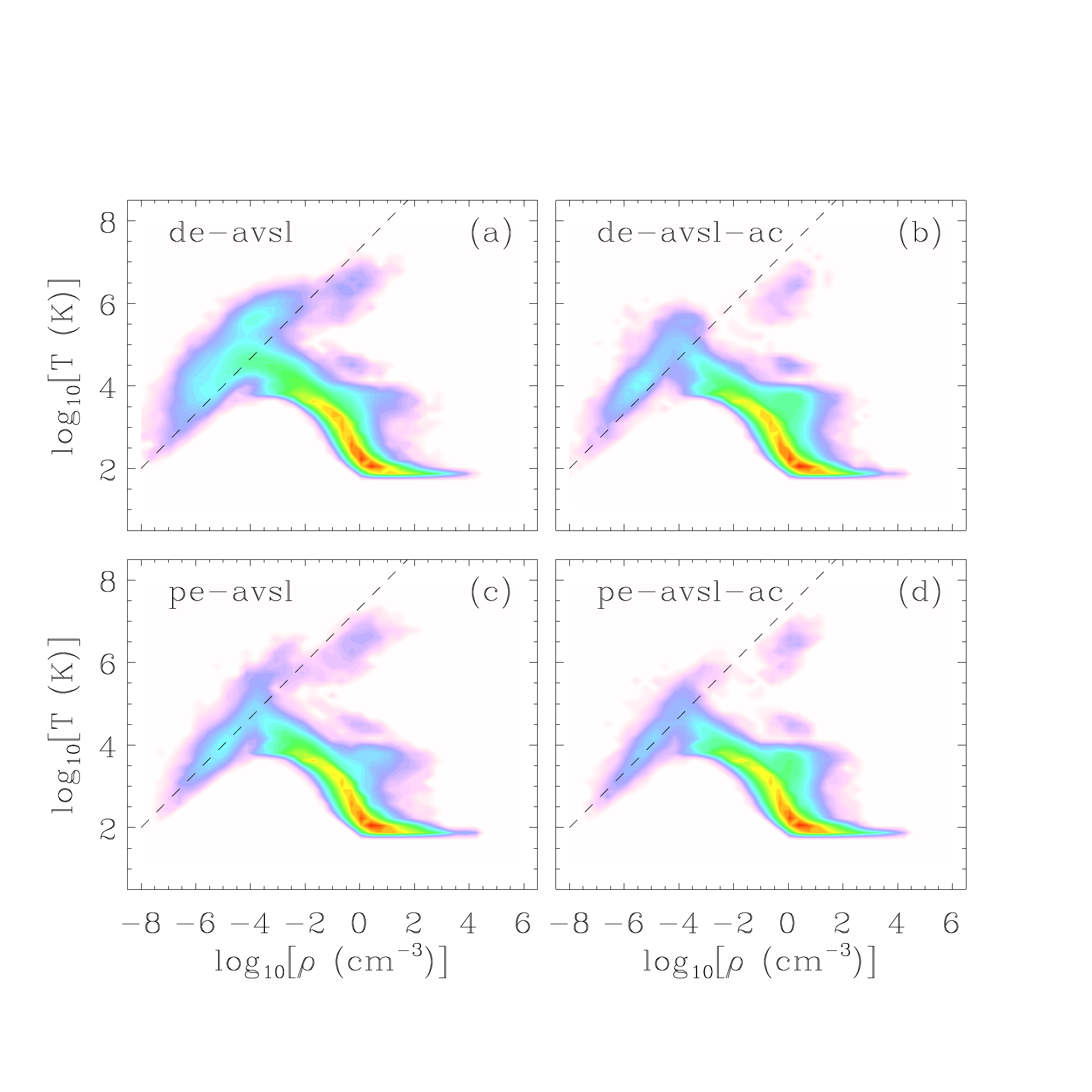}
	\caption{	
	The phase diagram of four different SPH schemes at t = 300 Myr. 
	{\it Panel (a):} DE formulation without AC. 
	{\it Panel (b):} DE formulation with AC. 
	{\it Panel (c):} PE formulation without AC.
	{\it Panel (d):} PE formulation with AC.
	The dashed line is the adiabatic relation $T \propto \rho^{\gamma-1}$. 
	The majority of the gas is cold ($< 10^4$ K) and follows a locus of thermal equilibrium where cooling balances heating. 
	The SN-induced hot gas ($\gtrsim 10^6$ K) on the disk is driving galactic outflows. 
	Outflowing gas, once being pushed out of the disk, has a long cooling time due to its low density and therefore follows an adiabatic relation. 
	Hot gas cools faster with more efficient fluid mixing and therefore the amount of outflow decreases.
	} 	
	\label{fig:pd_4in1}
\end{figure}

\textbf{
We show in Fig. {\ref{fig:pd_4in1}} the phase diagrams of four different SPH schemes ('de-avsl', 'de-avsl-ac', 'pe-avsl', and 'pe-avsl-ac') at t = 300 Myr.
The majority of the gas follows a locus of thermal equilibrium where the cooling balances heating.
The SN feedback injects both thermal energy to the gas which raises the temperature to $\gtrsim 10^6$ K at density $\sim$ 1 cm$^{-3}$ as well as kinetic energy which helps to dissolve the dense clumps.
The highly over-pressurized hot gas drives the ambient ISM into galactic outflows.
Once the gas is pushed out of the disk,
radiative cooling and heating becomes inefficient due to the low density,
and the temperature and density of the gas follows an adiabatic relation $T \propto \rho^{\gamma-1}$ (dashed lines).
Fluid mixing provides an extra channel for the hot gas to cool.
Therefore, a fraction of hot gas drops rapidly to low temperatures.
Indeed, for the models that includes AC (right panels),
we see more gas around $\rho \sim$ 10 cm$^{-3}$ and $T \sim 10^{3.5}$ K,
as a result of more efficient mixing.
There is also a small amount of gas at $\rho \sim$ 10 cm$^{-3}$ and $T \sim 10^{4.5}$ K in all cases, which is due to the local minimum of the cooling curve at that temperature.
More quantitatively,
the phase diagram can be divided into different phases
by defining 
the cold (hot) disk gas as $T \leqslant 10^5$ K ($T > 10^5$ K) and $|z| <$ 5kpc,
and the outflow/inflow gas as $|z| \geq$ 5kpc,
where z = 0 is the disk plane.
From panel (a) to (d) in Fig. {\ref{fig:pd_4in1}},
the mass fraction of the hot disk gas is $0.54\%, 0.22\%, 0.34\%$ and $0.17\%$,
while the mass fraction of the outflow/inflow gas is $5.16\%, 1.25\%, 1.3\%$ and $0.61\%$, respectively (the rest is the cold disk gas).
In our fiducial model (panel d) where mixing is the most efficient,
the mass fraction of cold (hot) disk gas is the largest (smallest), while the amount of outflow is almost an order of magnitude smaller than the 'de-avsl' model.
Since the hot gas in the disk cools faster due to more efficient fluid mixing,
the effect of feedback is weakened and the amount of outflow decreases.
We will discuss the outflow in more detail in the next section.
}

\subsection{Star formation and galactic outflows}\label{sec:SFR}
In this section we investigate the global star formation rate (SFR) as 
well as the outflow rate from the galactic disk for the 
different SPH schemes. We show in Fig. {\ref{fig:sfr_ml_MW_highZ}, panel (a), the time
evolution of SFR. Surprisingly, there is no obvious difference in
terms of SFR between the traditional constant AV and our variable AV, despite the
presence of the instability in the former.  This is because the SFR is
mainly determined by the total amount of the star forming gas and is
not too sensitive to the morphology of the gas. Including AC or using the PE
formulation both increase the overall SFR by about a factor of
two. This is a consequence of more efficient mixing between the
hot and cold gas, which leads to the presence of more cold
gas available for star formation, in agreement with \citet{2013MNRAS.428.2840H}.  
The gradually declining SFR is expected as the total amount of
star-forming gas is continuously depleted either by galactic outflows
or the star formation itself.

Fig. {\ref{fig:sfr_ml_MW_highZ}, panel (b), shows the mass loading $\eta$,
defined as the ratio of the outflow rate to the SFR.
The outflow rate is computed by summing the total mass of gas particles passing through a plane at z = $\pm$ 5 kpc, where the z-axis is the rotation axis of the disk. 
There is an initial burst of outflow in all cases caused by
the initial setup. Since all the gas particles are initially located
in the disk, the environment outside of the disk is a vacuum which
allows an unimpeded outflow triggered by the SN explosions. These
outflowing particles will be pulled back down to the potential well of the
disk due to gravity, leading to the onset of gas inflow which can
interact with the subsequent outflow. Eventually the inflow and
outflow reach a quasi-steady state and a hot gaseous halo is
formed (after $\sim$ 150 Myrs).
A small fraction of the gas can even escape the halo completely.
The mass loading $\eta$ shows a strong 
dependence on the adopted SPH schemes: both the PE formulation and the
AC suppress the mass loading. 
One direct explanation is that the
increased SFR transforms part of the gas into stars, thus reducing the
amount of gas available to the outflow. However, the mass loading
drops almost an order of magnitude between the two most extreme cases
while the difference of SFR is only about a factor of 2.  
This suggests that the feedback efficiency is directly affected by the SPH
implementation. As it becomes weaker (for AC and PE models) the hot
gas cannot be blown out of the disk as efficiently. 
The driving force of
the shock is dissipated as the shocked heated particles are
continuously mixed with the ambient cold ISM.
We also observe that the effect (reduced mass loading) seems stronger at higher resolution due to more efficient mixing.
This is contrary to \citet{2013arXiv1311.2073H} where they find both the SFR and mass loading are insensitive to the adopted SPH scheme.
The discrepancy might be due to the different feedback implementation:
in \citet{2013arXiv1311.2073H} the contribution of momentum input (the radiation pressure) is significant, 
which is expected to be less affected by fluid mixing.

\begin{center}
\begin{figure*}
\includegraphics[trim = 0mm 0mm 0mm 0mm, clip, width=6.5in]{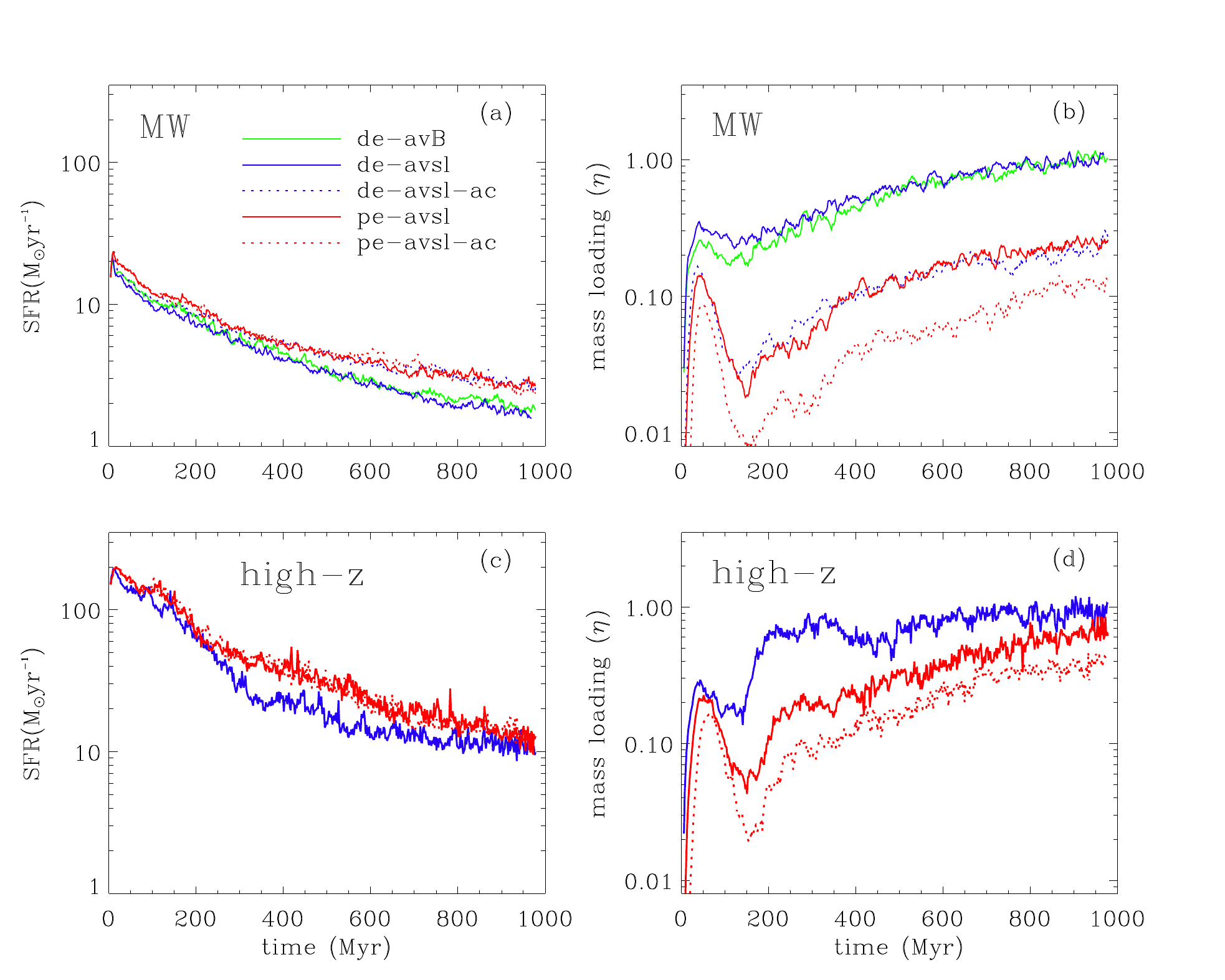}
\caption{The star formation rate and mass loading as a function of time for a Milky-Way like galaxy (panel (a), (b)) and a high-redshift galaxy (panel (c), (d)), using different SPH schemes: DE using constant viscosity plus the Balsara
          switch (de-avB, green line), DE  using variable
          strong limiter AV without (de-avsl, blue line) and with
          (de-avsl-ac, blue dashed line)
          artificial conduction, and PE using a variable AV without
          (pe-avsl, red line) and with (pe-avsl-ac, red dashed line) AC. 
          While the SFR shows only a weak dependence on the adopted SPH scheme, 
          the outflow rate (hence the mass loading) is much more sensitive 
          to the adopted SPH scheme. }
\label{fig:sfr_ml_MW_highZ}
\end{figure*}
\end{center}

We further investigate the star formation rate and the mass loading using a different initial condition which may represent a gas-rich high-redshift disk galaxy, set up by the same method as in Section \ref{sec:disk}.
The galaxy contains $6.4 \times 10^{10} \rm{M_{\odot}}$ of gas and $2.7 \times 10^{10} \rm{M_{\odot}}$ of stars in the disk, and a stellar bulge of $6.4 \times 10^{10} \rm{M_{\odot}}$. The dark matter halo follows a Hernquist profile with a concentration of 3.5 and a mass of $1.5 \times 10^{12} \rm{M_{\odot}}$. Again we set a radial metallicity gradient of -0.04 dex/kpc. 
The particle numbers for the different components are $N_{\rm halo} = 3\times 10^5$ for the dark matter halo, $N_{\rm disk} = 3\times 10^5$ for the stellar disk, $N_{\rm bulge} = 7\times 10^5$ for the stellar bulge and $N_{\rm gas} = 7\times 10^5$ for the gaseous disk.
The softening lengths are 100 pc for the dark matter halo and 20 pc for the gas, disk and bulge components.
Half of the gas follows an exponential profile with a scale length of 3 kpc, while the other forms an extended flat disk with a radius of 15 kpc.
The scale height of the stellar disk is 0.3 kpc and the scale length of the bulge is 1.2 kpc.
Fig. \ref{fig:sfr_ml_MW_highZ}, panel (c), shows the star formation rate with three different SPH schemes: 'de-avsl', 'pe-avsl', and 'pe-avsl-ac'. As in the Milky Way case, the SFR differs only slightly for different SPH schemes, though a systematic increase can still be seen for schemes that promote mixing. Interestingly, the mass loading in Fig. \ref{fig:sfr_ml_MW_highZ}, panel (d), becomes less sensitive to the adopted SPH scheme compared to the Milky Way case, which is in better agreement with \citet{2013arXiv1311.2073H}. The difference between the two extreme cases is only a factor of two after the quasi-steady state has reached ($\sim$ 400 Myr). One possible explanation is that the star formation rate is so high so that it drives outflow more easily despite the increased capability of fluid mixing.
From the test cases presented here - and for this specific feedback implementation - it seems that with the traditional SPH ('de-avsl') the outflow properties (mass loading of order unity) are independent of the initial conditions (star formation rate). 
On the other hand,
our fiducial model ('pe-avsl-ac') has a factor of 10 lower mas loading for the MW disk and only a factor of $\sim 3$ lower mass-loading for the high-z disk. Apparently this introduces a SFR dependent mass-loading. 
\textbf{
The outflow velocity also depends on the adopted SPH scheme.
The mean velocities projected on z-axis at z = $\pm$ 5 kpc,
after reaching a quasi-steady state,
are about 200, 140, and 100 km/s for the 'de-avsl', 'pe-avsl', and 'pe-avsl-ac' model, respectively.
This trend is consistent with the picture that fluid mixing suppresses outflows.
We refrain from a detailed comparison with observations as the main goal of this work is to investigate the effect of different SPH schemes.
}

\subsection{Accretion from the hot gaseous halo}\label{sec:accretion}
In this section we investigate the properties of hot halo gas that might be accreting onto the disk.
A hot gaseous halo is added to the Milky-Way like disk using the method presented in \citet{2011MNRAS.415.3750M}.
The density follows a $\beta$-profile:
\begin{equation}
	\rho_{\rm hg}(r) = \rho_0 \left[ 1 + \left( \frac{r}{r_{\rm c} } \right)^{2}\right]^{-1.5 \beta},
\end{equation}
where $\rho_0$ is the core density, $r_c$ is the core radius and $\beta$ describes the slope of the profile at large radii.
We set $r_c = 0.22 r_s$ and $\beta = 2/3$ as in \citet{2011MNRAS.415.3750M}.
The core density $\rho_c$ is determined by the total baryonic fraction $f_b$ within the virial radius $r_{\rm vir}$.
We set $f_b$ = 0.12 such that the mass of the hot gaseous halo within $r_{\rm vir}$ is $M_{\rm hg} = 6.54\times10^{10} \rm{M_{\odot}}$.
The temperature profile is determined assuming hydrostatic equilibrium at a given radius.
The hot gaseous halo is rotating along the rotation axis of the disk where the specific angular momentum of the gas is $\alpha$-times the the specific angular momentum of the dark matter halo. 
We set $\alpha$ = 4 following \citet{2011MNRAS.415.3750M} who found this choice to agree best with observational constraints.
The metallicity of the hot gaseous halo is set $Z = 0.3 ~Z_{\rm{\odot}}$ based on observations of Milky Way's hot gaseous halo \citep{2013ApJ...770..118M}.
The particle numbers for different components are $N_{\rm halo} = 6\times 10^5$ for dark matter halo, $N_{\rm disk} = 4.8\times 10^5$ for the stellar disk, $N_{\rm bulge} = 1.5\times 10^5$ for the stellar bulge, $N_{\rm cg} = 1.2\times 10^5$ for the cold gaseous disk, and $N_{\rm hg} \approx 1.3\times 10^6$ for the hot gaseous halo.
The individual particle mass is five times larger than in Section \ref{sec:holes} and \ref{sec:SFR}.
The softening lengths are 100 pc for the dark matter halo and 20 pc for the gas, disk and bulge components.

Fig. \ref{fig:accretion} shows the gas density at t = 1.8 Gyr using different SPH schemes in both face-on and edge-one views. The morphological difference in accretion properties is striking: 
in the 'de-avsl' case, the inflowing gas forms small blobs which are purely numerical artifacts due to the lack of proper fluid mixing.
Including AC efficiently eliminates these blobs, which is in agreement with \citet{2013MNRAS.434.1849H}.
The PE formulation ('pe-avsl') also prevents such blobs, while the accretion morphology is different from the 'de-avsl-ac' case; the latter is smoother and more filamentary.
This might reflect the difference of mixing mechanisms between AC and PE formulation.
The PE formulation allows mixing only in a dissipation-less way; mixing occurs via turbulent motion with entropy still conserved at particle level.
On the other hand, AC explicitly smooths the entropy gradient and therefore the morphology is expected to be smoother.
Our fiducial model combines both ('pe-avsl-ac') and shows also filamentary structures.
\textbf{
\citet{2013MNRAS.434.1849H} found that in their favored SPH scheme (SPHS) which avoids numerical blobs by AC, the accreting overdense filaments are able to fragment into clumps via nonlinear thermal instability triggered by the SN-driven outflows.
We do not find such fragmentation in out fiducial model probably because our feedback is too weak to induce enough nonlinear over-densities.
}

\begin{center}
\begin{figure*}
\includegraphics[trim = 0mm 0mm 0mm 0mm, clip, width=7.in]{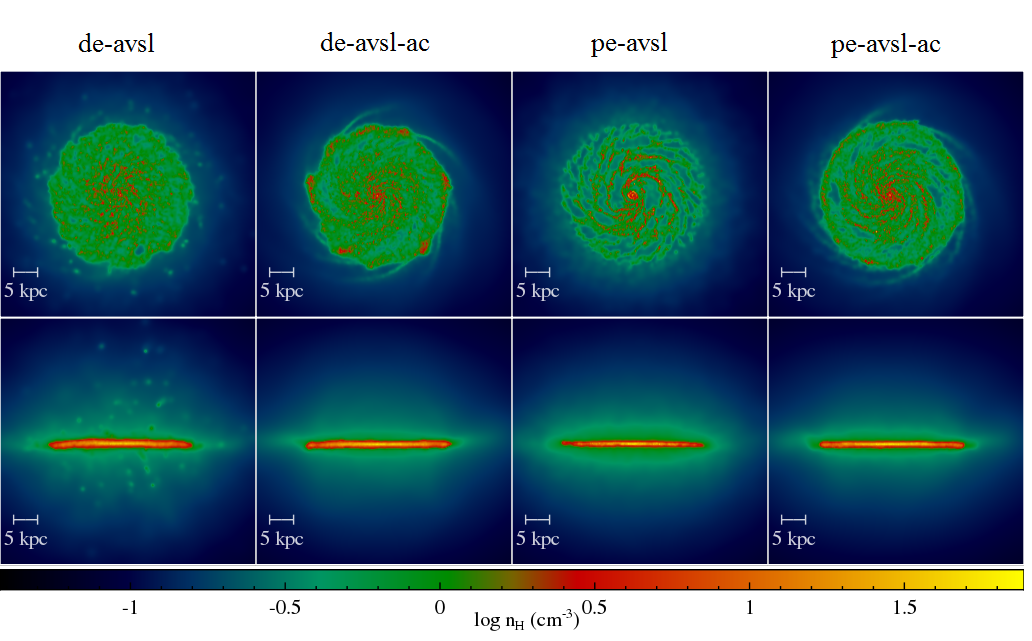}
\caption{The face-on (top row) and edge-on (bottom row) views of the projected gas density at $t$ = 1.8 Gyr, with four different SPH schemes. 
{\it First column}: DE formulation without AC (de-avsl). 
{\it Second column}: DE formulation with AC (de-avsl-ac).
{\it Third column}: PE formulation without AC (pe-avsl). 
{\it Fourth column}: PE formulation with AC (pe-avsl-ac). 
The 'de-avsl' scheme leads to blobs which are purely numerical artifacts due to the lack of proper fluid mixing. These problematic blobs can be efficiently avoided by either including AC or PE formulation (or both).
}
\label{fig:accretion}
\end{figure*}
\end{center}

\section{Summary \& Discussion}
\label{sec:conclusion}
In this paper we study the performance of different algorithmic
implementations of the SPH code GADGET for a variety of idealialized
hydrodynamic tests including the Gresho, Sod shock tube, Sedov
explosion, 'equilibrium square', Keplerian ring, Kelvin-Helmholtz, and
the 'blob' test (Section \ref{sec:hydrotest}). We also test the impact
on the dynamical evolution of 
a Milky Way like disk galaxy - including cooling from a hot gaseous halo - 
as well as a gas-rich high-redshift disk galaxy including metal cooling, metal enrichment, star formation and feedback from supernova explosions.

We test a density-entropy and a pressure-entropy formulation in
combination with recently proposed implementations for the treatment
of artificial viscosity and conduction (Section \ref{sec:method}). The
study indicates that implementations with a pressure-entropy
formulation \citep{2013ApJ...768...44S,2013MNRAS.428.2840H} in
combination with a Wendland $C^4$ kernel with 200 neighbors,
variable artificial viscosity with higher order shock detection
\citep{2010MNRAS.408..669C} and artificial conduction
\citep{2008JCoPh.22710040P, 2012MNRAS.422.3037R} pass all idealized hydrodynamic tests.      

We propose a modified implementation for artificial viscosity with a
strong viscosity coefficient limiter with high sensitivity to the
velocity curl (Section \ref{sec:AV}). In addition to passing all the
idealized tests, our fiducial implementation performs very well for
the disk evolution simulations, where some other implementations can
develop a viscous instability (Section \ref{sec:disk}). Despite the
successes our fiducial code still shows weaknesses in low Mach number
shear flows and the rate of convergence. Still, the above results
indicate improvements with respect to most standard SPH
implementations
(e.g. \citealp{2005MNRAS.364.1105S,2009ApJS..184..298W,2011A&A...529A..27H}).
\textbf{
These improvements, along with the advantages discussed in the introduction, 
keep SPH a competitive numerical method. 
Comparison between results of SPH and grid-based methods (e.g. \citealp{2002A&A...385..337T}; \citealp{2014ApJS..211...19B}) can also provide important information as they have very different numerical artifacts and are highly complementary.
}

The actual algorithmic implementation has a significant impact on
the gas outflow properties of our disk evolution models. 
Implementations with better fluid mixing capabilities in
general result in reduced mass loading. 
In the Milky Way case,
our fiducial implementation
has about an order of magnitude lower mass loading compared to a
standard implementation, while at the same time the star formation rate is only
slightly higher. 
For the gas-rich, high-redshift disk model,
the mass loading becomes less sensitive (but still notable) to the adopted SPH scheme.
In addition, we demonstrate that either artificial conduction or a pressure-entropy formulation (or both) suppresses the formation of artificial cold blobs. 
\textbf{
We note that the star formation model as well as the feedback recipe have also significant impact on the evolution of galaxies (e.g., \citealp{2010MNRAS.409.1541S}, \citealp{2012MNRAS.423.1726S}).
However,
an accurate scheme for the hydrodynamics is equally (if not more) important,
as the star formation and feedback models need to be calibrated under the 'correct' hydrodynamics.
The impact of star formation and feedback models will be addressed in future work.
}

In Table \ref{table:sum}, we present a summary of all tests in this
paper. SPH implementations marked with '$\checkmark$' passed the
respective test while those marked with '$\times$' failed. Again,
these tests indicate that SPH can perform well with a more accurate kernel, a
pressure-entropy formulation, artificial viscosity with a strong, curl
sensitive limiter and a higher order shock indicator and artificial
conduction.
Our fiducial implementation, which we term SPHGal, passes all tests. 
\footnote{ We find (not shown) that the weak limiter scheme 'pe-avwl-ac' also passes all of the idealized tests that might show differences with the strong limiter (i.e. shock tube, Sedov, Keplerian ring and the blob test). It only fails in the disk simulation, where both shocks and shear flows coexist.}

\begin{table}
\caption{Summary of the results of the idealized hydrodynamic tests in
  Section \ref{sec:hydrotest} and the galactic disk test in Section
  \ref{sec:disk}. SPH implementations marked with 
  $\checkmark$ passed the test while those marked with $\times$ failed the
  test. Our fiducial implementation 'pe-avsl-ac' passed all tests.}  
\label{table:sum}
\begin{tabular}{| l | l | l |}    
\hline\hline
Test&                 SPH scheme&                 Results      \\
\hline
Gresho&               de-avB (cs64)&                $\times$ \\
      &               de-avsl-ac (wld200)&          $\checkmark$   \\
      &               pe-avsl-ac (wld200)&          $\checkmark$    \\
\hline
Shock tube&           pe-avsl-ac&                   $\checkmark$     \\
          &           pe-avwl-ac&                   $\checkmark$     \\
\hline
Sedov     &           de-avsl&                      $\checkmark$     \\    
          &           pe-avsl&                      $\times$  \\
          &           de-avsl-ac    &               $\checkmark$     \\
          &           pe-avsl-ac    &               $\checkmark$     \\
          &           pe-avwl-ac    &               $\checkmark$     \\
          &           pe-avsl-ac-erho&              $\times$  \\
\hline
Keplerian ring &      de-avB-lvg       &            $\times$  \\
               &      pe-avsl-ac-lvg   &            $\times$  \\
               &      de-avB           &            $\times$  \\
               &      pe-avsl-ac       &            $\checkmark$    \\
               &      pe-avwl-ac       &            $\checkmark$    \\
\hline
Hydrostatic    &      de-avsl-ac       &            $\times$  \\
               &      pe-avsl          &            $\checkmark$     \\
               &      pe-avsl-ac       &            $\checkmark$     \\
\hline 
Kelvin-Helmholtz  &   pe-avsl-ac       &            $\checkmark$     \\
\hline            
Blob       &   pe-avsl-ac       &            $\checkmark$     \\
           &   pe-avwl-ac       &            $\checkmark$     \\
\hline
Galactic disk &           de-avB&                      $\times$     \\    
          &           de-avwl&                      $\times$  \\ 
          &           de-avwl-ac    &               $\times$     \\
          &           de-avsl    &               $\checkmark$     \\
          &           de-avsl-ac &              $\checkmark$  \\
          &           pe-avB&                      $\checkmark$     \\    
          &           pe-avwl&                      $\times$  \\  
          &           pe-avwl-ac    &               $\times$     \\
          &           pe-avsl    &               $\checkmark$     \\
          &           pe-avsl-ac &              $\checkmark$  \\
\end{tabular}
\end{table}

In the follwing we discuss our main results in more detail: 

\begin{itemize}
\item
With the PE formulation the spurious surface tension at contact
discontinuities is eliminated by construction and fluid mixing can be
modeled properly without the help of AC 
\citep{2010MNRAS.405.1513R,2013ApJ...768...44S, 2013MNRAS.428.2840H}. 
We therefore suppress AC in
shear flows to prevent over-mixing by including a quadratic limiter
similar to the Balsara switch (Section \ref{sec:ac}).  Another advantage of
this limiter is that it avoids unwanted conduction in a
self-gravitating system in hydrostatic equilibrium.

\item
It is necessary to include the AC in the PE formulation if strong
shocks are involved (Section \ref{sec:sedov}). The weakness of the PE
formulation stems from the entropy-weighted sum in the pressure
estimate, which makes the results noisy and even biased if the entropy
variation is large from one particle to another. Including AC greatly
improves the results in the Sedov explosion test. In addition, the
entropy-weighted density gives 
a biased estimate (the pre-shock 'bump') in the Sedov explosion test
even when the particles are distributed regularly. The mass-weighted
density is a more reliable estimate and should be used whenever the
information of density is  needed.  
Indeed, the only entropy-weighted
quantity we use is the pressure estimate in the equation of motion.
Using the entropy-weighted density in the dissipation terms leads to an
incorrect prediction of the shock position.

\item
We find a feedback-driven instability developing in a typical
isolated disk galaxy if the gaseous disk becomes too viscous.  The hot
bubbles created by SN explosions are too viscous to be sheared away by
the differential rotation.  They merge with one another and eventually
form large holes in the disk.  The AV switch with a strong limiter is
able to suppress the viscosity so that the disk remains stable.  On the other
hand, it could also be possible that it suppresses viscosity too much
and cannot capture shocks properly.  However, the Sedov explosion test
is recovered equally well and provides credibility for our
modified AV scheme.  In the cases where both shocks and shear flows
coexist, it is in fact difficult to determine how much viscosity is
appropriate. Compromises have to be made between properly modeling
shocks and avoiding artificial shear viscosity. This seems to be a
general issue for most SPH schemes that use AV for shock capturing (an
exception might be 'Godunov SPH', e.g. \citealp{2010MNRAS.403.1165C},
\citealp{2011MNRAS.417..136M}). Adopting the PE formulation and 
including AC alleviates the situation, which suggests that
spurious surface tension also plays a role for the instability. The
boundaries between the hot bubbles and the cold ISM are sustained by
the spurious surface tension in the DE formulation. As such, they are
difficult to destroy once being created. 

\item
We have investigated the star formation rate as well as the mass loading
of the disk galaxy models with different SPH schemes.
The SFR increases slightly when we adopt the SPH schemes that allow
more efficient fluid mixing. Here the hot gas in the disk (created by
SN explosions) is continuously mixed with the large amount of cold gas 
and therefore cools faster.  For this reason, the amount of cold gas
available to star formation is higher.  
The mass loading is affected even more and drops by almost an order of
magnitude (in the MW case) from a standard SPH implementation to our fiducial
implementation. This is partly related to the increased SFR that
transforms some of the gas into stars.  In addition, the continuous
mixing with the ambient ISM weakens the 
driving force of the shock-heated particles and hence decreases the
mass loading. This is in contrast
to \citet{2013arXiv1311.2073H} who have made similar improvements 
to their SPH scheme and claim that they in general find little difference in
SFR as well as the mass loading using different SPH schemes. 
\textbf{
The discrepancy may arise from the different feedback models.
It is expected that the thermal feedback would be more sensitive to the fluid mixing, as mixing provides an extra channel for the hot gas to cool and thus weakens the feedback.
However, we note that the kinetic feedback would still convert some fraction of kinetic energy into thermal energy via shock heating and therefore is not completely unaffetced by mixing.
}
In the case of the gas-rich, high-redshift disk model, the mass loading becomes less sensitive to the adopted SPH scheme, which is in better agreement with \citet{2013arXiv1311.2073H}.
One possible explanation is that the higher star formation rate drives the outflow more efficiently despite the increased capability of fluid mixing and hence weakening the differences.

\item
We investigate the accretion behavior of the hot gas halo onto the Milky Way like disk. 
As commonly being criticized in the literature,
the traditional SPH (DE without conduction) generates plenty of small blobs 
due to the lack of proper fluid mixing.
Including AC efficiently eliminates these blobs,
in agreement with \citet{2013MNRAS.434.1849H}.
The PE formulation, with or without AC, 
is also capable of eliminating the blobs.
In addition,
depending on the presence of AC,
the accretion morphology shows notable difference:
inflowing gas forms smooth filamentary structures if AC is included.
This might originate from the different mixing mechanism between AC and PE formulation.
In PE formulation, mixing is a dissipation-less, entropy-conserving process that relies on turbulent motion.
Hot and cold gas do not mix if the turbulent motion is weak.
On the other hand, AC promotes mixing by explicitly smoothing out entropy differences, 
leading to smoother morphology.

\end{itemize}

In summary, we have presented an updated SPH implementation SPHGal that
performs more accurately in several idealized hydrodynamic tests. The
outstanding problem in the traditional SPH, i.e. the poor capability
of fluid mixing, no longer exists thanks to the PE formulation.  In
addition, we have a controllable diffusion mechanism (the artificial
conduction) similar to the implicit numerical diffusion in the
grid-based methods. The higher order velocity gradients prevent false
triggering of AV. The AV switch, which involves a strong limiter, is
able to capture shocks properly and also to avoid too much viscosity
in a disk galaxy. The convergence rate is significantly improved
(still slower than the grid-based methods) and shows no sign of
saturation with increasing resolution, though the intrinsic issue
remains when we move further into the subsonic regimes. 
We conclude that, with all the modifications, SPHGal is an accurate and valuable 
numerical method for galaxy formation simulations and many other astrophysical applications.

\section*{Acknowledgements}
We would like to thank Claudio Dalla Vecchia and Michael Aumer for their collaboration, generous support
and most 
valuable discussions.  We would also like to thank Walter Dehnen, Volker Springel,
Justin Read, Daniel Price 
and James Wadsley for critical and valuable comments. Many of the ideas presented
here were discussed 
at the MPA SPH workshop in February 2013, where the idea for this paper was born.   
We have used SPLASH \citep{2007PASA...24..159P} for the visualization. TN acknowledges support by
the DFG cluster of 
excellence 'Origin and Structure of the Universe'. TN and SW acknowledge support
from the DFG priority program 
SPP 1573 'Physics of the interstellar medium'. LO acknowledges support from the MPA
visitor programme.

\bibliographystyle{mn2e}
\bibliography{ref}

\appendix
\section{High Order Estimate of The Velocity Gradient} \label{app:hvg}
While we use a low order estimate of the pressure gradient in the equation of motion (so as to keep the exact conservation), we are free to use a high order estimate for the velocity gradient without any trade-off in accuracy. Here, we follow the approach of \citet{2012JCoPh.231..759P}, which is equivalent to \citet{2010MNRAS.408..669C} though with a slightly different derivation. The Greek letters ($\alpha, \beta, \gamma$) stand for the coordinate index and the Roman letters are particle labels.

The commonly used estimate of the velocity gradient is
\begin{equation}\label{eq:common_gv}
	(\widehat{\nabla\otimes{\bf v}})_{\alpha\beta} = 
	\frac{1}{\rho_i} \sum_{j} m_j ({\bf v}_j-{\bf v}_i)^{\beta}\nabla_i^{\alpha} W_{ij}.
\end{equation}
Expanding $v_j^{\beta}$ around $i$:
\begin{equation}
	v_j^{\beta} = v_i^{\beta} + \partial_{\gamma} v_i^{\beta} ({\bf x}_j - {\bf x}_i)^{\gamma} + O(h^2).
\end{equation}
Substituting into (\ref{eq:common_gv}) leads to 
\begin{equation}
	\sum_{j} m_j ({\bf v}_j-{\bf v}_i)^{\beta}\nabla_i^{\alpha}W_{ij} = 
	\partial_{\gamma} v_i^{\beta} \sum_{j} m_j ({\bf x}_j - {\bf x}_i)^{\gamma} \nabla_i^{\alpha}W_{ij}.
\end{equation}
We can obtain the improved estimate $\partial_{\gamma} v_i^{\beta}$ from a matrix inversion $\bf{X} = \bf{M}^{-1} \bf{Y}$ where ${\bf X}_{\gamma\beta} \equiv \partial_{\gamma} v_i^{\beta} $ and
\begin{eqnarray}	
\begin{aligned}
	{\bf M}_{\alpha\gamma}
	&\equiv\sum_{j} m_j ({\bf x}_j - {\bf x}_i)^{\gamma} \nabla_i^{\alpha}W_{ij} \\	
	&= \sum_{j} m_j ({\bf x}_j - {\bf x}_i)^{\gamma} ({\bf x}_i - {\bf x}_j)^{\alpha}
	\frac{1}{x_{ij}}\frac{\partial{W_{ij}}}{\partial{x_{ij}}} , 
\end{aligned}\\ \
\begin{aligned}
	{\bf Y}_{\alpha\beta} 
	&\equiv \sum_{j} m_j ({\bf v}_j-{\bf v}_i)^{\beta}\nabla_i^{\alpha}W_{ij} \\
	&= \sum_{j} m_j ({\bf v}_j-{\bf v}_i)^{\beta} ({\bf x}_i - {\bf x}_j)^{\alpha}
	\frac{1}{x_{ij}}\frac{\partial{W_{ij}}}{\partial{x_{ij}}} .
\end{aligned}
\end{eqnarray}
The velocity divergence, shear tensor, and vorticity can be obtained readily from the velocity gradient:
\begin{equation}
	\nabla\cdot{\bf v} = \partial_{\alpha} v^{\alpha} 
\end{equation}
\begin{equation}
	{\bf S}_{\alpha\beta} = \frac{1}{2} (\partial_{\alpha} v^{\beta} + \partial_{\beta} v^{\alpha}) 
	- \frac{1}{3}\nabla\cdot{\bf v}\delta_{\alpha\beta}
\end{equation}
\begin{equation}
	(\nabla\times{\bf v})_{\gamma} = \epsilon_{\alpha\beta\gamma}\partial_{\alpha} v^{\beta} .
\end{equation}

\end{document}